 \documentclass[aps, twocolumn, showpacs, amsmath]{revtex4}
\usepackage{dcolumn}% Align table columns on decimal point
\usepackage{bm}% bold math
\usepackage{graphicx}% Include figure files
\usepackage{color}
\usepackage{subfigure}
\usepackage{hyperref}
\usepackage{latexsym}
\usepackage{amsthm}
\usepackage{amssymb}

\DeclareGraphicsExtensions{.jpg,.pdf, .mps, .png, .eps, .ps, .EPS,.gif}

\begin{document}
\def\be{\begin{equation}}
\def\ee{\end{equation}}

\def\bc{\begin{center}}
\def\ec{\end{center}}
\def\bea{\begin{eqnarray}}
\def\eea{\end{eqnarray}}
\newcommand{\avg}[1]{\langle{#1}\rangle}
\newcommand{\Avg}[1]{\left\langle{#1}\right\rangle}

\def\ie{\textit{i.e.}}
\def\etal{\textit{et al.}}
\def\m{\vec{m}}
\def\G{\mathcal{G}}

\newcommand{\gin}[1]{{\bf\color{magenta}#1}}
\newcommand{\bob}[1]{{\bf\color{red}#1}}
\newcommand{\bobz}[1]{{\bf\color{magenta}#1}}

\title{Topological Percolation on  Hyperbolic Simplicial Complexes}

\author{Ginestra Bianconi}
\affiliation{The Alan Turing Institute, 96 Euston Rd, London NW1 2DB, United Kingdom\\ 
School of Mathematical Sciences, Queen Mary University of London, London, E1 4NS, United Kingdom}
\author{Robert M. Ziff}
\affiliation{Center for the Study of Complex Systems and Department of Chemical Engineering, University of Michigan, Ann Arbor, Michigan 48109-2136, USA}
\begin{abstract} 
Simplicial complexes are increasingly used to understand the topology of complex systems as different as brain networks and social interactions. It is therefore of special interest to extend the study of percolation to simplicial complexes. Here we propose a topological theory of percolation for discrete hyperbolic simplicial complexes. Specifically we consider hyperbolic manifolds  in dimension $d=2$ and $d=3$ formed by simplicial complexes,  and we investigate their percolation properties in the presence of topological damage, i.e.,  when nodes, links, triangles or tetrahedra are randomly removed. We  show that in $d=2$  simplicial complexes there are four  topological percolation problems and in $d=3$, there are six.   We demonstrate the presence of two percolation phase transitions characteristic of hyperbolic spaces for the different variants of topological percolation. While most of the known results on percolation in hyperbolic manifolds are in $d=2$, here we  uncover the rich critical behavior of  $d=3$ hyperbolic manifolds, and show that triangle percolation displays a Berezinskii-Kosterlitz-Thouless (BKT) transition.  Finally we provide evidence that topological percolation can display a critical behavior that is unexpected if  only  node and link percolation are considered.
\end{abstract}


\maketitle

\section{Introduction}

Simplicial complexes  uncover the topology and geometry of interacting systems such as brain networks \cite{Bassett,Vaccarino2,Blue_Brain}, granular materials \cite{Granular,Nanoparticles} and social interaction networks \cite{Petri_Barrat}.
Simplicial complexes are able to capture higher-order interactions that cannot be encoded in a simple network. In fact they are not just formed by nodes and links but also by higher dimensional simplices {such as triangles,} tetrahedra and so on. 

Simplicial complexes, built by geometrical building blocks, are natural objects to study network geometry \cite{Emergent,Doro_manifold}. In particular hyperbolic simplicial complexes  \cite{CQNM,NGF,Hyperbolic,Polytopes} reveal emergent functionalities of complex networks \cite{Ana} and provide a major avenue to explore the very active  area of network hyperbolicity \cite{Kleinberg,Boguna1,Boguna2,Gromov1,Gromov2}.  

Percolation theory \cite{Doro_book,crit,Ziff_review,Kahng_review} studies the properties of the connected components when nodes or links are damaged with probability $q=1-p$, fully capturing the network robustness to failure events. However in simplicial complexes, topological damage can be directed not just to nodes and links but also to higher dimensional simplices.  In this respect a major question is how  to characterize the robustness of simplicial complexes to topological damage. Given the large variety of systems that can be described by simplicial complexes,  this is a  challenging  theoretical problem   of primary importance also for applications.
Here we  explore the robustness of hyperbolic simplicial complexes by introducing the framework of {\it topological percolation} and find that the response to  damage of higher dimensional simplices can be unexpected if one considers exclusively node or link percolation. 
 
Percolation theory \cite{Doro_book,crit,Ziff_review,Kahng_review} of complex networks has been widely studied in the past twenty years. While in uncorrelated random networks percolation is well known to give a single, continuous second-order  phase transition in hierarchical  networks, it is possible to observe a discontinuous phase transition \cite{hyperbolic_Ziff} or  a Berezinkii-Kosterlitz-Thouless (BKT) \cite{B,KT} transition \cite{flower_tau,BKT2,renormalization}. 
Additionally a  BKT percolation transition  is found also on percolation on growing networks \cite{Anomalous,Newman,Kahng2018}. Finally, there is growing interest in investigating generalized percolation problems such as explosive percolation \cite{dSouza,Ziff_explosive,Doro_explosive} and percolation on interdependent multiplex networks   \cite{Bianconi_book,Havlin,Grassberger,Doro_multiplex} that have been recently shown to display  anomalous critical behavior.

On hyperbolic networks, percolation theory has been shown to display not one but two percolation transitions   \cite{Schramm}  at the so-called lower $p^{l}$ and upper $p^{u}$ percolation thresholds. Below the lower percolation threshold (for $p<p^{l}$) there is no infinite cluster, above the upper percolation threshold (for $p>p^{u}$) an extensive infinite cluster exists, and for $p^{l}<p<p^{u}$ the average size of the largest cluster is infinite but sub-extensive.
Interestingly it has been shown \cite{hyperbolic_Ziff}, using a renormalization approach, that for hyperbolic $d=2$ dimensional manifolds called Farey graphs \cite{Farey0,Farey}  the transition at the upper critical dimension is discontinuous.
Several works have {studied} percolation  \cite{Patchy,clusters,Ziff_sierpinski,percolation_Apollonian,flower_tau,BKT2,renormalization} and the Ising and Potts models \cite{Boettcher_Ising,Boettcher_Potts} on other hierarchical networks, finding  both continuous {and discontinous phase transitions}. However most of the results on percolation in hyperbolic networks \cite{Gu_Ziff,Moore_Mertens} are restricted to $d=2$ spaces.
Here we explore how the scenario change in dimension $d=3$ and we address the general question whether the nature of the transition changes with the dimension of the manifold.
We consider a class of ``holographic"  hyperbolic simplicial complexes  in $d=2$ and $d=3$  that can be extended naturally in higher dimensions. If one only focuses on the nodes and links these simplicial complexes reduce to Farey graphs \cite{Farey0,Farey} and to Apollonian networks \cite{apollonian1} for dimension $d=2$ and $d=3$ respectively.

To study the robustness of these simplicial complexes  we propose the general framework of topological percolation that includes for each network several percolation problems and  is able to capture for each simplicial complex its response to different types of topological damage.  
Topological percolation  {expands} on previously known types of percolation transitions (node, link and $k$-clique) percolation. In fact while in node {and} link percolation random damage is  directed either to the nodes or to  the links  of a network, in topological percolation damage can be directed also to higher-dimensional simplices like triangles, tetrahedra and so on.
{ 
The topological percolation problems in higher dimensions are very closely connected to $k$-clique percolation \cite{clique1,clique2} or equivalently (using the topology term) $k$-connectedness \cite{Kahle}  where two $k$-cliques are considered connected if they share a $(k-1)$-clique.  However  the topological $k$-connectedness as well as  $k$-clique percolation have been  studied only for nondamaged networks while topological percolation consider the effects of topological damage. }

Topological percolation for $d=2$ simplicial complexes reduces to four percolation problems and for $d=3$ simplicial complexes reduces to six percolation problems.
Topological percolation on the considered hyperbolic simplicial complexes is naturally studied on generalized line graphs which take the form either of single {or}  multiplex networks \cite{Bianconi_book}.
Taking advantage of these line graphs here we show that topological percolation {in} both the $d=2$ and $d=3$ dimensional hyperbolic manifolds under consideration displays in general two percolation thresholds (except the trivial case of link percolation on the $d=3$ hyperbolic manifold).
We show however that the nature of the phase transition at the upper percolation threshold can change significantly. In particular our investigation of triangle percolation in the $d=3$ hyperbolic manifold displays a BKT transition not observed for any of the topological percolation problems in the Farey simplicial complex.

This paper is structured as follows. In Sec.\ II we define topological percolation on simplicial complexes, in Sec.\ III we characterize the hyperbolic manifolds under consideration in this work, in Sec.\ IV we define the general percolation properties of hyperbolic manifolds, in Sec.\ V we discuss topological percolation in the $d=2$ hyperbolic manifold, in Sec.\ V study topological percolation in the $d=3$ hyperbolic network, in Sec.\ VI we compare topological percolation for $d=2$ and $d=3$ hyperbolic manifolds. Finally in Sec.\ VII we give the conclusions.

\begin{figure*}
    \includegraphics[width=1.99\columnwidth]{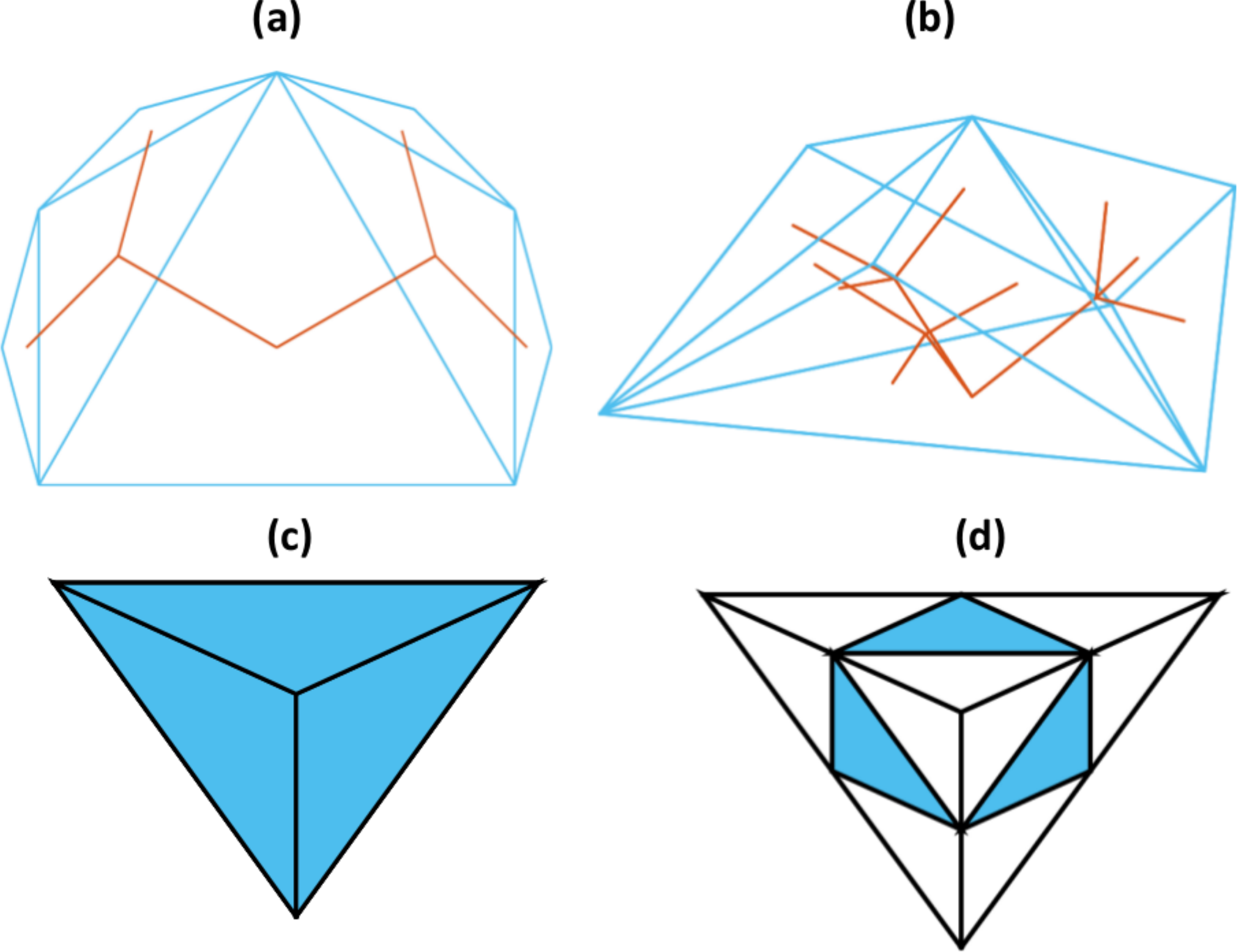}
	\caption{ Panel (a) shows the $d = 2$ skeleton of {the} Farey simplicial complex  at iteration $n=3$ skeleton (in blue)   together with the Cayley tree constituting its generalized line graph (in red). Panel (b) shows the $d=3$ hyperbolic manifold skeleton (in blue) at iteration $n=2$  together with the Cayley tree constituting one of its generalized line graphs (in red). Panel (c) shows the Apollonian network at iteration $n=1$. Panel (d) shows that the Sierpinski gasket simplicial complex (in blue) is a generalized line graph of the Apollonian network (both graphs are shown at iteration $n=1$). }
	\label{fig1}
\end{figure*}
\begin{figure}
    \includegraphics[width=0.99\columnwidth]{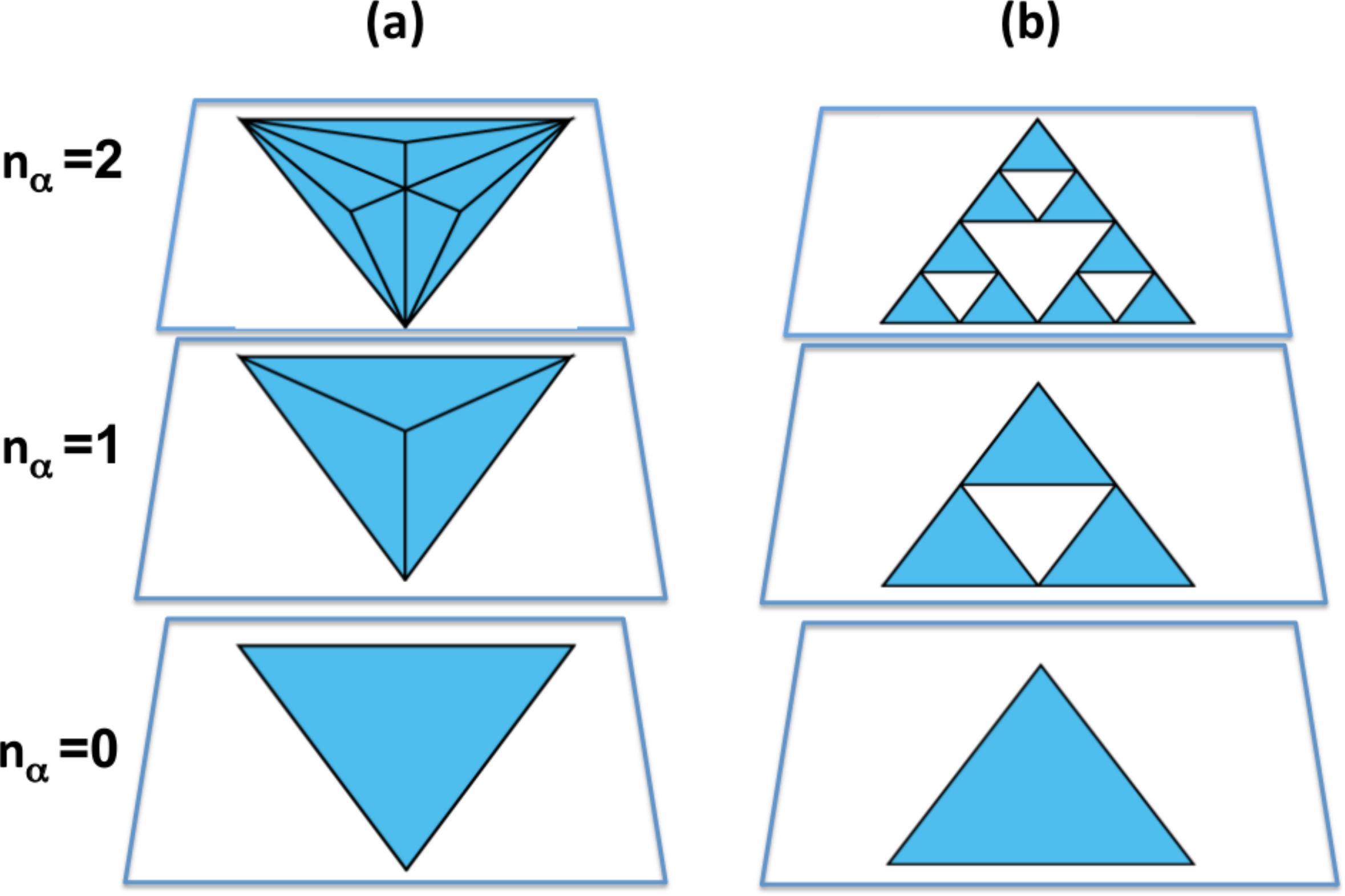}
	\caption{Panel (a) shows the multiplex network description of the $d=3$ hyperbolic manifold.  Panel (b) show the multiplex Sierpinski gasket that describes a generalized line graph of the $d=3$ hyperbolic manifold.}
	\label{fig2}
\end{figure}

\section{Topological percolation on hyperbolic simplicial complexes}

Simplicial complexes are not just formed by nodes ($0$-simplices), and links ($1$-simplices) but also by higher-dimensional simplices such as triangles ($2$-simplices), tetrahedra  ($3$-simplices) and so on. Simplicial complexes are the natural objects used in topological data analysis and they are also particularly useful to investigate network geometry because {they} are formed by geometrical building blocks.
Let us now give a more formal definition of simplices and simplicial complexes. 

A {\em simplex} of dimension $d$ ($d$-simplex) is formed by a set of $d+1$ nodes. The $\delta$-faces of a $d$-simplex $\mu$ are the $\delta$-simplices (with $\delta<d$) that can be constructed by taking a subset of $\delta+1$ of the nodes of $\mu$.
A {\em simplicial complex} ${\mathcal K}$ of dimension $d$ is formed by a set of $\delta$-dimensional simplices with $\delta\leq d$  glued along their faces  (where $d$ is the maximum of all dimensions $\delta$ of the simplices belonging to $\mathcal K$) with the two properties:
\begin{itemize}
\item[(a)] if the simplex $\mu$ belongs to the simplicial complex (i.e., $\mu \in {\mathcal K}$) all its faces $\mu'\subset \mu$ also belong to the simplicial complex (i.e., $\mu' \in {\mathcal K}$).
\item[(b)] if two simplexes $\mu,\mu'$ belong to the simplicial complex (i.e., $\mu \in {\mathcal K}$ and $\mu' \in {\mathcal K}$) their intersection is either empty (i.e., $\mu\cap \mu'=\emptyset$)  or belongs to the simplicial complex (i.e., $\mu\cap \mu' \in \mathcal K$).
\end{itemize}  
Note that the  {\em skeleton} of a simplicial complex is the network formed by its nodes and links.

{ Simplicial complexes are related  to hypergraphs \cite{Zlatic}---in fact they can   be both considered as  sets of sets of nodes. However simplicial complexes satisfy both condition (a) and (b) not satisfied by the hypergraphs. Simplicial complexes  are therefore  closed  under the operation of taking non-empty subsets of each set and this allows for the study of topological properties as a function of the dimensions of its simplices \cite{Kahle}.}

Here we introduce  topological percolation which extends and generalizes node and  link percolation to simplicial complexes.
In node percolation the properties of the connected components are monitored as a function of the probability $q=1-p$ to remove nodes, in link percolation  the same properties are studied when links are removed with probability $q$.

{ In network science stardard percolation has been leveraged by the notion of $k$-clique percolation \cite{clique1,clique2} called also by the topologists $k$-connectedness \cite{Kahle} where cliques of nodes are connected if they share $(k-1)$-cliques. Therefore $k$-connected clusters are formed by the  sets of $k$-connected  $k$-cliques. However so far the properties of the $k$-connected clusters have not been studied in the presence of  damage.}

 Here we consider  {\em topological percolation} that studies the properties of the   $k$-connected clusters \cite{clique1,clique2,Kahle} of the simplicial complexes in presence of {\em topological damage} that is not only directed to nodes or links but also to higher dimensional simplices.

The choice of the term {\em topological percolation} is due to the fact that in topology it is usually the case that the properties of a network (or of a graph) are leveraged to the properties of higher dimensional simplices. For instance the graph Laplacian is leveraged to the concept of higher-order Laplacian \cite{Jost}. 
Similarly topological percolation includes standard node and link percolation but leverage these two problems also to higher dimensions.
Moreover topological percolation characterizes the connectivity of a simplicial complex in presence of topological damage that can produce nontrivial Betti numbers (``holes") in the simplicial complex \cite{Perspectives,Lambiotte}.   

In topological percolation we   consider percolation among $\delta$-dimensional faces (with $\delta<d$) of the simplicial complex connected through $(\delta+1)$-faces removed with probability $q$.
For instance in  topological percolation in $d=3$  hyperbolic manifolds includes the distinct percolation problems:
\begin{itemize}
\item {\it Link percolation (also known as bond percolation).}  In this case nodes are connected to nodes through links that are removed with probability $q$.  
\item {\it Triangle percolation.} In this case links are connected to links through triangles that are removed with probability $q$.
\item {\it Tetrahedron percolation.} In this case triangles are connected to triangles through tetrahedra that are removed with probability $q$.  
\end{itemize}
Moreover topological percolation includes  the percolation among $\delta$-dimensional faces (with $0<\delta<d$) of the simplicial complex connected through $(\delta-1)$-faces removed with probability $q$.
For simplicial complexes in $d=3$ therefore topological percolation includes also the distinct percolation problems: 
\begin{itemize}
\item {\it Node percolation (also known as site percolation).} In this case links are connected to links through nodes that are removed with probability $q$.  
\item {\it Upper-link percolation.} In this case triangles are connected to triangles through links (at the edges of the triangles)  that are removed with probability $q$.
\item  {\it Upper-triangle percolation.} In this case tetrahedra are connected to tetrahedra through {common} triangles that are removed with probability $q$.
 \end{itemize}
 Therefore for $d=3$ simplicial complexes topological percolation includes six percolation problems.
 For $d=2$ simplicial complexes clearly there are only four percolation problems including link percolation, triangle percolation, node percolation and upper-link percolation.
In general topological percolation on a $d$-dimensional simplicial complex includes $2d$ percolation problems. 

{ Note that in node percolation it is {customary} to indicate that if a node is removed all its links are also removed while for link percolation only the links are removed. This difference is actually redundant if
one {focuses} exclusively on the property of the connectivity of the simplices, as in node percolation a  link can be connected to a link  only through a node so if we remove that node all the links connected to it will not be part of the percolating cluster automatically. Therefore our definition of the first class of topological percolation problems (including link percolation) is symmetric to the definition of the second set of problems (including node percolation) and there is no need of a distinction between the two cases.}

\section{Hyperbolic networks under consideration}

Here we study topological percolation on  two classical examples of hyperbolic lattices in dimension $d=2$ and $d=3$ and we describe them using simplicial complexes.
The first simplicial complex under consideration is the  Farey simplicial complex (whose skeleton is the Farey graph \cite{Farey0,Farey}). This is an infinite $d=2$ hyperbolic simplicial complex  constructed iteratively starting from a single link (see Fig.\ $\ref{fig1}$a).  
At iteration $n=1$ we attach a triangle to the initial link.
At iteration $n>1$ we attach a triangle to every link to which we have not yet attached a triangle.
{The number of nodes $N_n^{(0)}$,  links $N_n^{(1)}$ and triangles $N_n^{(2)}$ at iteration $n$ are given by} 
\bea
N_n^{(0)}&=&1+2^{n},\nonumber \\
N_n^{(1)}&=&2^{n+1}-1,\nonumber \\
N_n^{(2)}&=&2^{n}-1.
\eea
The second  simplicial complex under consideration is the  $d=3$ hyperbolic simplicial complex that   generalizes  the Farey simplicial complex in  dimension $d=3$.
This is an infinite $3$-dimensional hyperbolic lattice {constructed} iteratively starting from a single triangle (see Fig.\ $\ref{fig1}$b).
At iteration $n=1$ we attach a tetrahedron to the initial triangle.
At iteration $n>1$ we attach a tetrahedron to each triangle to which we have not yet attached a tetrahedron.
{The number of nodes $N_n^{(0)}$,  links $N_n^{(1)}$, triangles $N_n^{(2)}$ and tetrahedra $N_n^{(3)}$ at iteration $n$ are given by }
\bea 
N_n^{(0)}&=&(5+3^{n})/2,\nonumber \\
N_n^{(1)}&=&(3+3^{n+1})/2,\nonumber \\
N_n^{(2)}&=&(3^{n+1}-1)/2,\nonumber \\
N_n^{(3)}&=&(3^{n}-1)/2.
\eea

If one focuses exclusively on its skeleton the $d=3$ hyperbolic simplicial complex we consider in this paper reduces to the Apollonian network \cite{apollonian1}, so at the network level the two are equivalent. The Apollonian network \cite{apollonian1} is a planar network constructed iteratively according to the following algorithm. At iteration $n=0$ the Apollonian network is formed by a single triangle. At each iteration $n>1$ each triangle in the $d=2$ plane is tessellated into three triangles by inserting a node in its center and connecting each of its nodes to the central node (see Fig.\ $\ref{fig1}$c). Note however that even though this construction generates the same network skeleton of the $d=3$ hyperbolic simplicial complex described above, the two models differ if one {considers} simplices of dimension $d=2$  (triangles) because in the planar description the triangles are naturally defined exclusively as the faces of the planar representation of the network and they are therefore a subset of the triangles included in the $d=3$ hyperbolic manifold. { Moreover while the $d=3$ manifold contains tetrahedra, the Apollonian network does not. }

Interestingly  the Apollonian network is closely related to the Sierpinski gasket \cite{Ziff_sierpinski}. In fact if we construct a network whose nodes correspond to the links of the Apollonian network and two nodes are connected if the corresponding links share a triangle, we obtain the Sierpinski gasket (see Fig.\ $\ref{fig1}$d).

{ We note here that the considered hyperbolic manifolds have ``holographic" properties \cite{Hyperbolic,NGF,Polytopes,Ana}. In fact   if we define the boundary as the set of all $(d-1)$-faces arrived at the last generation and all their faces, we observe that the considered hyperbolic manifolds  in dimension $d=2$ and dimension $d=3$ have all the nodes at the boundary.} Moreover also all the links are at the boundary therefore   at the network level no node or link is lost if we consider the projection of the network on the boundary of the simplicial complex. It follows that the  network skeleton of the hyperbolic manifold in dimension $d=2$ reduces  to a hierarchical $d=1$ network \cite{hyperbolic_Ziff} and the one of the hyperbolic manifold in dimension $d=3$  reduces to the planar ($d=2$) Apollonian network \cite{apollonian1}.  This is a peculiar property of these hyperbolic structures shared with the models of emergent geometry in Refs.\ \cite{CQNM,NGF,Hyperbolic,Polytopes,Ana} but not shared by other hyperbolic manifolds  with a bulk such as the ones studied in Refs. \cite{Gu_Ziff, Moore_Mertens}. 

To uncover the equations for topological percolation in these lattices it is opportune to define {a} suitable generalization of line graphs  to simplicial complexes.
A line graph of a network is constructed by placing a node for each link of the original network and connecting these nodes if the corresponding two links are connected by a node in the original network.

Line graphs can be clearly extended to higher dimensions when studying simplicial complexes.
For instance in Fig.\ $\ref{fig1}$a we {show} a network whose nodes correspond to the triangles of the Farey simplicial complex and whose links connect nodes corresponding to adjacent triangles in the Farey simplicial complex. Interestingly this network is a Cayley tree of coordination $z=3$ and will be particularly useful as a reference to study triangle percolation and upper-link percolation on the Farey simplicial complex.
For the  $d=3$ hyperbolic network it is possible to construct a Cayley tree of coordination $z=4$ whose nodes correspond to tetrahedra and links connect nodes corresponding to adjacent tetrahedra (see Fig.\ $\ref{fig1}$b).
This Cayley graph will be particularly useful to study tetrahedron percolation and upper-triangle percolation.
The generalized line graph whose nodes are links of the $d=3$ hyperbolic manifold and nodes are connected if the corresponding links share a triangle in the original hyperbolic network is a multiplex network with $n+1$ number of layers in which {each} layer  is formed by  a Sierpinski gasket  (for a definition of multiplex network see Ref.\ \cite{Bianconi_book}).
To show this let us observe that the $d=3$ hyperbolic manifold at iteration $n$ can be  described as a multiplex network  of $n+1$ layers where each layer $n_{\alpha}=0,1,\ldots, n$ is an Apollonian network at iteration $n_{\alpha}$ (see Fig.\  $\ref{fig2}$a). This construction allows {one} to take explicitly  into account the simultaneous presence of triangles entering the $d=3$ manifold at each iteration.
Using the relation between the Apollonian network and the Sierpinski gasket  discussed above it is natural to realize that the generalized line graph of this multiplex {network} whose nodes are the links of the $d=3$ manifold and two nodes are connected if the corresponding links of the $d=3$ manifold are incident to the same triangle, can be described as a multiplex Sierpinski gasket of $n+1$ layers. In this multiplex Sierpinski gasket  each layer $n_{\alpha}=0,1,\ldots, n$ is a Sierpinski gasket at iteration $n_{\alpha}$ (see Fig.\  $\ref{fig2}$b). 
This is the structure  on which triangle percolation can be naturally studied. 

{ We note that the different topological percolation problems have very different natures. For instance both link percolation and upper link percolation investigate the effect of damage directed to links; however, while the first one is the traditional link percolation problem the second one studies the  $3$-connected clusters when links are randomly removed. However  all the topological percolation problems can be mapped into node or link percolation in a suitably chosen generalized line graph. For instance  upper-link percolation can be recast into a node percolation problem in the generalized line graph in which the links of the original simplicial complex are the nodes and nodes are connected if the corresponding links of the original simplicial complex belong to the same triangle.}

\begin{table}
\center
\caption{\label{table1} Lower $p^{l}$ and upper $p^{u}$ percolation thresholds for topological percolation on the $d=2$ and $d=3$ hyperbolic manifolds under consideration. {The section of the paper in which each percolation problem is treated is also indicated.}}
\footnotesize
\begin{tabular}{|llll|}
\hline
$d=2$ &section&$p^{l}$ & $p^{u}$\\
\hline
Link percolation &(A1) & $0$ &$\frac{1}{2}$ \\
\hline
Triangle percolation &(A2)&$\frac{1}{2}$ &$1$\\
\hline
Node percolation & (A3)&$0$ & $1$\\
\hline
Upper-link percolation &(A4)&$\frac{1}{2}$&$1$\\
\hline
\hline
$d=3$ &section&$p^{l}$&$p^{u}$\\
\hline
Link percolation &(B1)&N/A &$0$ \\
\hline
Triangle percolation & (B2)&$0$ & $0.307981\ldots$\\
\hline
Tetrahedron percolation&(B3)&$\frac{1}{3}$&$1$\\
\hline
Node percolation &(B4)& $0$ & $1$\\
\hline
Upper-link percolation & (B5)&$0$ & $1$\\
\hline
Upper-triangle percolation& (B6)&$\frac{1}{3}$ &$1$\\
\hline
\end{tabular}
\end{table}
\section{General properties of topological percolation on the studied hyperbolic manifolds}

Node and link percolation on hyperbolic manifolds \cite{Schramm} {  and in general non-amenable graphs \cite{Lyon,Hasegawa} are known} to have not just one {but} two percolation thresholds: the lower $p^{l}$ and  the upper $p^{u}$ percolation thresholds. 
In particular it is found that the phase diagram of percolation include three regions.
\begin{itemize}
\item
For $p<p^{l}$ no cluster has infinite size.
\item
For $p^{l}<p<p^{u}$ the network has an infinite but sub-extensive maximum cluster of  average size $R$
\bea
R\sim N^{\psi}
\eea
with $0<\psi<1$.
Here $N$ indicates the number of nodes of the network and $\psi$ is called the {\em fractal critical exponent}.
\item
For $p>p^{u}$ the network has an extensive cluster, i.e., the fraction $M$ of nodes in the giant component scales like
\bea
M\simeq \frac{R}{N}=O(1).
\eea
\end{itemize}
Here we {find} that these general properties of node and link percolation on hyperbolic lattices remain valid also for the higher-dimensional problems for topological percolation on simplicial complexes (see Table $\ref{table1}$).
However we {find} that the value of the thresholds, the critical fractal exponent, and the nature of the transition can change significantly for the different versions of the topological percolation and with the overall dimension $d$ of the manifold as will be detailed in the next sections.

\section{Topological Percolation on $d=2$ Hyperbolic Manifold}

In this section we will consider topological percolation on the $d=2$ Farey simplicial complex in detail.
We will summarize known results on link percolation \cite{hyperbolic_Ziff} and we will show the critical behavior of node, triangle and upper-triangle percolation.

\subsection*{(A1)  Link percolation} In link percolation,  links  are removed with probability $q$ and the connected component are formed by nodes connected to nodes through intact links.
This transition in the $d=2$ Farey simplicial complex has been studied  by Boettcher, Singh and Ziff in Ref.\ \cite{hyperbolic_Ziff}.\\  

The probability $\hat{T}_{n+1}$ that the two nodes which appeared in the simplicial complex at iteration $n=0$ are connected at the generation $n+1$ is given by  \cite{hyperbolic_Ziff}
\bea
\hat{T}_{n+1}=p+(1-p)\hat{T}_n^2.
\eea
In fact they are either directly connected (event which occurs with probability $p$) or if they are not directly connected (event which occurs with probability $q=1-p$), they can be connected if each  node  is connected to the node arrived in  the network at iteration $n=1$ (event which occur with probability $\hat{T}_n^2$).
In the limit $n\to \infty$ this equation has the steady-state solution  
\bea
\hat{T}_{\infty}=\left\{\begin{array}{ccc}\frac{p}{1-p}&\mbox{for} &p<\frac{1}{2}\nonumber \\1&\mbox{for} &p\geq\frac{1}{2}\end{array}\right..
\eea
Since we have $\hat{T}_{\infty}>0$ for any $p>0$ and $\hat{T}_{\infty}=1$ for $p\geq \frac{1}{2}$,  the lower $p_1^{l}$ and the upper $p_1^{u}$ critical thresholds are given by    
\bea
p^{l}_1=0,\ \ \ p^{u}_1=\frac{1}{2}. 
\eea
To investigate the nature of the phase transition, Boettcher, Singh and Ziff in Ref.\ \cite{hyperbolic_Ziff} have proposed a theoretical approach based on the generating functions $T_{n}(x)$ and $S_n(x,y)$. In a Farey simplicial complex at iteration $n$ the  function $T_n(x)$ is the generating function of the  
number of nodes in the connected component linked to both initial nodes. The function $S_n(x,y)$  is the generating function for the sizes of the two connected components linked  exclusively to one of the two initial nodes. 
These generating functions are given by 
\bea
T_n(x)&=&\sum_{\ell=0}^{\infty}t_n(\ell) x^{\ell},\nonumber \\
S_n(x,y)&=&\sum_{\ell,\bar{\ell}}s_n(\ell,\bar{\ell})x^{\ell}y^{\bar{\ell}}.
\eea
Here we consider  the  $d=2$ hyperbolic manifold at iteration $n$, and we indicate with $t_n(\ell)$  the distribution of the number of nodes $\ell$ connected to the two initial nodes and with  $s_n(\ell,\bar{\ell})$ we indicate  the joint distribution of  the number of nodes $\ell$ connected exclusively to a given initial node and the number of nodes $\bar{\ell}$ connected exclusively to the other initial node.

The recursive equations for $T_n(x)$ and $S_n(x,y)$ start from the initial condition $T_0(x)=p$ and $S_0(x,y)=1-p$ and read \cite{hyperbolic_Ziff}
\begin{widetext}
\bea
T_{n+1}(x)&=&p\left\{xT_n^2(x)+2xT_n(x)S_n(x,x)+S_n(1,x)S_n(1,x)\right\}+(1-p)xT_n^2(x)\nonumber \\
S_{n+1}(x,y)&=&(1-p)\left\{xT_n(x)S_n(x,y)+yS_n(x,y)T_n(y)+S_n(1,x)S_n(1,y)\right\}.
\label{T2xlink}
\eea
\end{widetext}
with 
\bea
\hat{T}_n=T_n(1)=1-S_n(1,1).
\eea
The size $R_n$ of the connected component linked to the initial  two nodes at iteration $n$ is given by 
\bea
R_n=\left.\frac{dT_n(x)}{dx}\right|_{x=1}.
\eea
By explicitly deriving $R_n$ from Eq.\ $(\ref{T2xlink})$ in Ref.\ \cite{hyperbolic_Ziff} it has been proven that  for $n\gg1 $, $R_n$ scales like
\bea
R_n\sim [N_n^{(0)}]^{\psi}
\eea
with
\bea
\psi=\frac{1}{\ln 2}\ln\left[\frac{1+3p-4p^2}{2q}+\sqrt{\frac{1-pq^2}{4q}}\right]
\eea
{where $q = 1 - p$,} for $p\leq \frac{1}{2}$. Moreover in Ref.\ \cite{hyperbolic_Ziff} it is also found 
 that 
\bea
M_{\infty}=\lim_{n\to \infty}\frac{R_{n}}{N_n^{(0)}}
\eea
has a discontinuous transition at $p^{u}_1=\frac{1}{2}$ with $M_{\infty}^{u}=0.609793\ldots$.
Interestingly this model is also very closely related to percolation in one-dimensional lattices with long-range links \cite{AN,Chayes}.

\subsection*{(A2) Triangle percolation } In triangle percolation, triangles are removed with probability     $q$ and the connected components are formed by links that  are connected to links through intact triangles. 
This is node percolation on the  Cayley tree network of degree $z=3$ where nodes are triangles and links connect two adjacent triangles. \\
We evaluate the percolation properties of this network by measuring the average number of triangles $R_n$ that at iteration $n$ are connected to the triangle added at iteration $n=1$.
At iteration $n=1$ we have clearly  $R_1=p$. For any  $n\geq 1$ $R_{n+1}$ is given by the recursive equation
\bea
R_{n+1}=p(z-1)R_{n},
\eea
{for arbitrary coordination number $z$} with explicit solution
\bea
R_n=[p(z-1)]^{n-1}R_1.
\label{RnTriangle}
\eea
Asymptotically, for $n\gg 1 $ we $R_n$ scales like
\bea
R_n\sim [N_n^{(2)}]^{\psi},
\label{scalingt2}
\eea
where 
\bea
\psi=\frac{\ln[p(z-1)]}{\ln 2}=\frac{\ln[2p]}{\ln 2}.
\label{psit2}
\eea
{for $z = 3$.}
Therefore there  are two percolation transitions at  
\bea
p_2^{l}=\frac{1}{z-1}=\frac{1}{2},\ p_2^{u}=1.
\eea
The lower percolation threshold $p_2^{l}$  is found by imposing $\psi=0$ and the upper percolation threshold $p_2^{u}$ is found by imposing $\psi=1$.
We  {denote the}  fraction of triangles  in the largest connected component in an infinite Farey simplicial complex { as $M_{\infty}$, defined by 
\bea
M_{\infty}=\lim_{n\to \infty} \frac{R_n}{N_n^{(2)}}.
\label{Minft2}
\eea
This order parameter  has a discontinuous transition at $p_2^{u}$ with 
\bea
M_{\infty}&=&\left\{\begin{array}{ccc}0 &\mbox{if} &p<p_2^{u} \\
1 &\mbox{if}&p=p_2^{u}\end{array}\right..
\label{MA2}
\eea

\subsection*{(A3)  Node percolation} In node percolation nodes are removed with probability $q$ and the connected components are formed by links  connected to links through intact nodes.\\
In order to study this percolation problem we consider a Farey simplicial complex at iteration $n$ and  we calculate the average number of nodes $R_n^{[++]}$ and $R^{[+-]}_n$ connected to the initial link given that its  end nodes are either both  intact  (case $[++]$) or one intact and one removed (case $[+-]$).
Starting from the initial condition $R^{[++]}_0=R^{[+-]}_0=0$ the values of $R^{[++]}_n $ and $R^{[+-]}_n$ are found by iteration of the equations,
\bea
{{\bf R}}_{n+1}={\bf B}{{\bf R}}_n+p{{\bf 1}}
\label{B}
\eea
where 
\bea
{{\bf R}}_n=\left(\begin{array}{c}{R}_{n}^{[++]}\\{R}_{n}^{[+-]}\end{array}\right), \  \ {{\bf 1}}=\left(\begin{array}{c}1\\1\end{array}\right).
\eea
The matrix ${\bf B}$ is given by
\bea
{\bf B}=\left(\begin{array}{cc}2p& 2(1-p)\\p&1\end{array}\right)
\eea
and has maximum eigenvalue 
\bea
\lambda=\frac{1}{2}+p+\frac{1}{2}\sqrt{1+4p-4p^2}.
\eea
The solution of Eq.\  $(\ref{B})$ is 
\bea
{\bf R}_n=p\sum_{r=0}^{n-1}{\bf B}^r {{\bf 1}}
\eea
Therefore the leading term for ${{\bf R}}_n$ is 
\bea
{{\bf R}}_n\sim p\lambda^n\sim [N_n^{(0)}]^{\psi}
\eea
with 
\bea
\psi=\frac{\ln \lambda}{\ln 2}.
\eea
By imposing $\psi=0$ we get $p_3^{l}$, {and} by imposing $\psi=1$  we get $p_3^{u}$, whose numerical values are
\bea
p_3^{l}=0,\ p_3^{u}=1.
\eea
At $p_3^{u}$ the fraction $M_{\infty}$ of nodes in the largest component of an infinite Farey simplicial complex, defined as
\bea
M_{\infty}=\lim_{n\to \infty}\frac{R_n}{N_n^{(0)}}
\eea 
has a discontinuous jump, i.e.,
\bea
M_{\infty}&=&\left\{\begin{array}{ccc}0 &\mbox{if} &p<p_3^{u} \\
1 &\mbox{if}&p=p_3^{u}\end{array}\right..
\label{MA3}
\eea

\subsection*{(A4) Upper-link percolation} In upper-link percolation links are removed with probability $q$ and one considers the connected components formed by triangles connected to triangles through intact links. 
This is link percolation on the  Cayley tree network of degree $z=3$ where nodes are triangles and links connect two adjacent triangles. \\
By indicating with $R_n$ the average number of triangles connected to the initial link at iteration $n$ we have $R_1=p$ and 
\bea
R_n=p(1-z)R_{n-1}.
\label{RnUpperLink}
\eea
These are the same equations found in triangle percolation  {(\ref{RnTriangle})}. Therefore we find 
that the lower and upper percolation thresholds are given by 
\bea
p_4^{l}=\frac{1}{2}, \ p_4^{u}=1
\eea
 with the same critical behavior found for triangle percolation. Therefore $R_n$ for $n\gg1 $ obeys the scaling in Eq.\  $(\ref{scalingt2})$ with  the fractal critical exponent $\psi$ given by Eq.\  (\ref{psit2}).
 Moreover at $p_4^{u}$ the order parameter $M_{\infty}$ defined as in Eq.\  (\ref{Minft2}) has a discontinuous jump described by 
\bea
M_{\infty}&=&\left\{\begin{array}{ccc}0 &\mbox{if} &p<p_4^{u} \\
1 &\mbox{if}&p=p_4^{u}\end{array}\right..
\label{MA4}
\eea

\section{Topological Percolation on $d=3$ hyperbolic manifold}

{ The characterization of all the topological percolation problems for the $d=2$ Farey simplicial complex has shown the ubiquitous presence of two percolation transitions typical of hyperbolic networks and a consistent presence of a discontinuous phase transition at $p=p^{u}$.}
Here our aim is to explore how this critical behavior {extends to} the $d=3$ hyperbolic simplicial complex under consideration.
Although the construction of the $d=3$ manifolds is a obvious extension of the construction of the  $d=2$ Farey simplicial complex, we need to {note} that the degree distribution of the Farey graph is exponential while the degree distribution of the Apollonian graph is power-law.
Interestingly this major difference of the skeleton of the two simplicial complexes under consideration is responsible for the differences in link percolation {between} these two structures.
In fact in the Apollonian network link percolation \cite{percolation_Apollonian} has a single continuous percolation threshold $p^{u}=0$ with a critical scaling characteristic of scale-free networks. 
However the other topological percolation problems behave very differently for this $d=3$ hyperbolic manifold. 
In particular here we show that all the other five topological percolation problems show the ubiquitous presence  of two percolation thresholds. Moreover the nature of the transition at the upper percolation threshold $p^{u}$ can vary. 
Particularly interesting is the study of  triangle percolation that here is shown to display a continuous  BKT percolation transition at the upper percolation threshold $p^{u}$. This critical behavior  is not observed at the level of node and link percolation. Therefore the response to topological damage can be unpredictable if only node and link percolation are considered.
\\

\subsection*{(B1) Link percolation} 
Given the equivalence between the skeleton of the $d=3$ hyperbolic simplicial complex and the Apollonian network, link percolation on the $d=3$ hyperbolic manifold reduces to link percolation in the  Apollonian network. This percolation problem has been  studied in Refs.\ \cite{percolation_Apollonian} and  \cite{Ziff_sierpinski}.
This is a particular case in which 
$p_1^{u}=0$ and therefore $p_1^{l}$ is not defined.
The percolation transition at $p=p_1^{u}$ is continuous \cite{percolation_Apollonian} and close to this phase transition, for   $p\ll 1$  the fraction of nodes $R_{\infty}$ in the giant component of an infinite network obeys  the scaling \cite{percolation_Apollonian}
\bea
R_{\infty}\sim e^{-c/p}
\label{MB1}
\eea
where $c>0$ is a constant.
Therefore in this case the fractal critical exponent  is $\psi=1$ for every $p>0$.

\begin{figure*}[ht!]   
    \includegraphics[width=1.20\columnwidth]{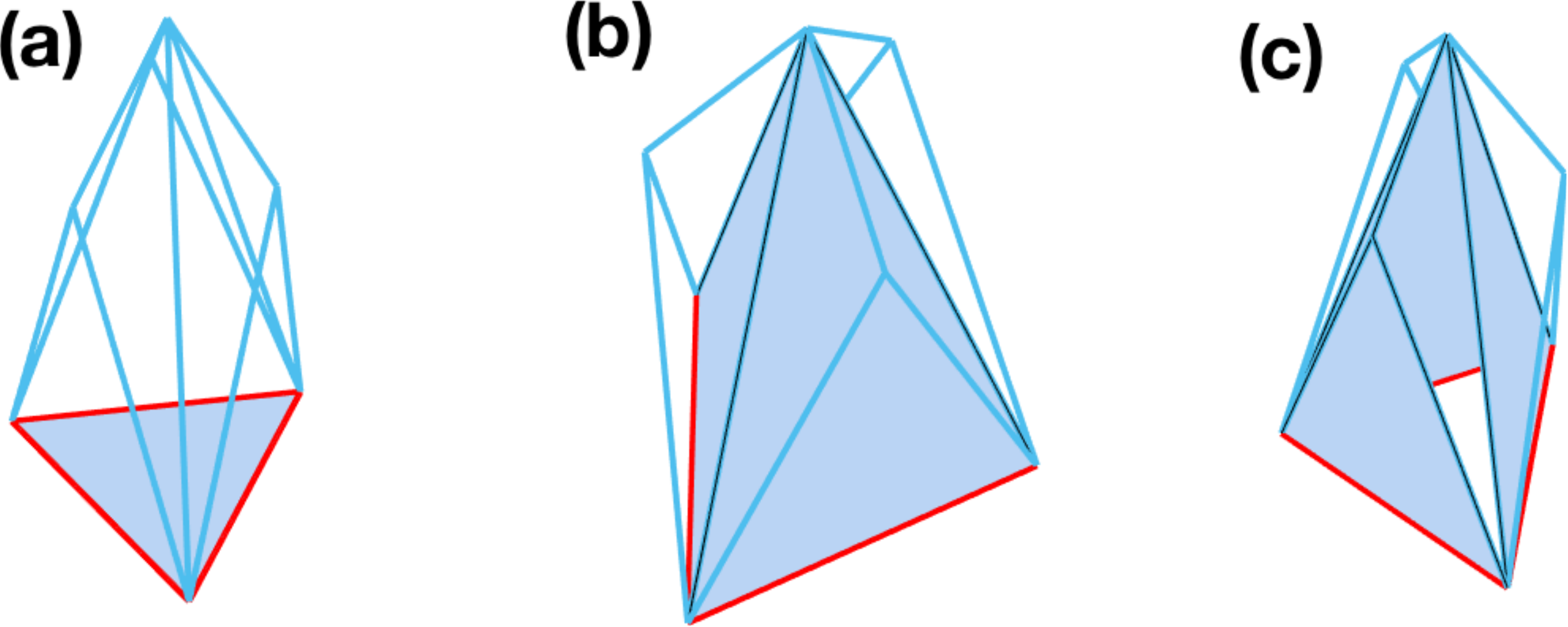}
	\caption{Examples of how  the three initial links of the $d=3$ hyperbolic manifold   {(indicated in red)} can be connected in triangle percolation: (panel a) through the presence of the initial triangle, (panel b,c) in absence of the initial triangle, through the presence of triangles entered at later iterations.}
	\label{fig3}
\end{figure*}

\subsection*{(B2) Triangle percolation} In triangle percolation triangles are removed with probability $q$ and the connected components are formed by links connected to links through intact triangles.
For instance the initial three links of the triangle at iteration $n=0$ can be connected if the initial triangle is not removed or, if it is removed, if they are connected through triangles added at a later iteration (see Fig.\  $\ref{fig3}$).

In order to study triangle percolation on the $d=3$ hyperbolic manifold we consider three variables $\hat{T}_n,\hat{S}_n,\hat{W}_n$ indicating, at iteration $n$, the probability that the three initial links are connected, that only two of the initial links are connected and that none of the initial links are connected  {respectively}.
In terms of the  multiplex Apollonian network with $n+1$ layers ($n$ iterations),
$\hat{T}_n$ is the probability that each pair of the  three original  links is connected (through triangles) in at least in one layer.
$\hat{S}_n$ is the probability that two links are connected in at least one layer (through  triangles) and the other link is not connected to the other two in any layer.
$\hat{W}_n$ is the probability that  no pair of links among the three initial nodes is connected in any layer.
The recursive equations for the probabilities $\hat{T}_n,\hat{S}_n,$ and $\hat{W}_n$ start from the initial condition $\hat{T}_0=p, \ \hat{S}_0=0$ and $\hat{W}_0=1-p$ and read
\begin{widetext}
\bea
\hat{T}_{n+1}&=&p+(1-p)(\hat{T}_n^3+6\hat{T}_n^2\hat{S}_n+3\hat{T}_n\hat{S}_n^2),\nonumber \\
\hat{S}_{n+1}&=&(1-p)\left[\hat{T}_n^2(\hat{S}_n+\hat{W}_n)+\hat{T}_n\hat{S}_n(7\hat{S}_n+2\hat{W}_n)+\hat{S}_n^2(4\hat{S}_n+\hat{W}_n)\right],\nonumber \\
\hat{W}_{n+1}&=&1-3\hat{S}_{n+1}-\hat{T}_{n+1}.
\label{Tri}
\eea
\end{widetext}
The first equation for $\hat{T}_{n+1}$ indicates that the three links of the original network are connected if the triangle that connects the three links is there in the layer $n_{\alpha}=0$ of the multiplex Apollonian network, and if the triangle is not there they must be connected through a chain of triangles in the layers $n_{\alpha}>0$. The second equation for $\hat{S}_{n+1}$ indicates that out of the three links of the original network only two are connected by  paths passing  through triangles.   This can occur only if the triangle connecting the three links directly in the layer $n_{\alpha}=0$ of the multiplex Apollonian network does not exist and two given  links are connected by triangles in the layer $n_{\alpha}>0$.  Finally the last equation is a normalization condition.
By using the vector $\hat{\bf V}_n=(\hat{T}_{n},\hat{S}_{n},\hat{W}_{n})$, the Eqs.\ $(\ref{Tri})$ can be written as 
\bea
\hat{\bf{V}}_{n+1}={\bf G}(\hat{\bf V}_n),
\eea
admitting the asymptotic solution $\hat{\bf V}_{\infty}$ for $n\to \infty$ satisfying
\bea
\hat{\bf{V}}_{\infty}={\bf G}(\hat{\bf V}_{\infty}).
\label{hatV}
\eea
This equation admits a solution $\hat{T}_{\infty}>0$ for every $p>0$ so the lower critical threshold is $p_2^{l}=0$.
Moreover this equation has a singular discontinuity at the point $p^{u}$ where the maximum eigenvalue $\Lambda_G$ of the Jacobian of ${\bf G}$ satisfies 
\bea
\left.\Lambda_{G}\right|_{\hat{\bf V}=\hat{\bf V}_{\infty}}=1.
\label{LambdaG}
\eea
In fact for $p>p^{u}$ the only stable solution is $\hat{T}_{\infty}=1,\hat{S}_{\infty}=\hat{W}_{\infty}=0$.
By imposing Eq.\  ($\ref{hatV}$) and $(\ref{LambdaG})$ we get that 
$p^{u}_2=0.307981 \ldots$
and $\hat{\bf V}_{\infty}^{u}=(0.509801,$ $0.0934843,$ $0.209745)$.  
Therefore the lower and upper critical thresholds are 
\bea
p_2^{l}=0, \ p_2^{u}=0.307981 \ldots.
\eea
The emergence of the discontinuity in $\hat{T}_{\infty}$ can be clearly appreciated from Fig.\  $\ref{fig7}$. For details on the numerical calculation see Supplementary Material \cite{SI}.

To investigate the nature of the upper percolation transition we define the generating functions $T_n(x),S_n(x,y)$ and $W_n(x,y,z)$.
These generating functions are defined as 
\bea
T_n(x)&=&\sum_{\ell=0}^{\infty}t_n(\ell) x^{\ell},\nonumber \\
S_n(x,y)&=&\sum_{\ell,\bar{\ell}}s_n(\ell,\bar{\ell})x^{\ell}y^{\bar{\ell}}\nonumber \\
W_n(x,y,z)&=&\sum_{\ell,\bar{\ell},\hat{\ell}}w_n(\ell,\bar{\ell},\hat{\ell})x^{\ell}y^{\bar{\ell}}z^{\hat{\ell}},
\eea
with 
\bea
\hat{T}_n=T_n(1),\ \hat{S}_n=1-S_n(1,1), \ \hat{W}_n=W_n(1,1,1).
\eea
Given a $d=3$ hyperbolic manifold at iteration $n$, $t_n(\ell)$ indicates the distribution of the number of links $\ell$ connected to the three initial links; $s_n(\ell,\bar{\ell})$ indicates the joint distribution of  the number of links $\ell$ connected  to two initial links and the number of links $\bar{\ell}$ connected exclusively   to the third initial link;
finally  $w_n(\ell,\bar{\ell},\hat{\ell})$ indicates the joint distribution of observing $\ell, \bar{\ell},\hat{\ell}$ links connected to each of the initial links. 
Note that given the definition above $W(x,y,z)$ is left unchanged by a permutation of its variables, while $S_n(x,y)$ is not.

\begin{figure}
    \includegraphics[width=0.90\columnwidth]{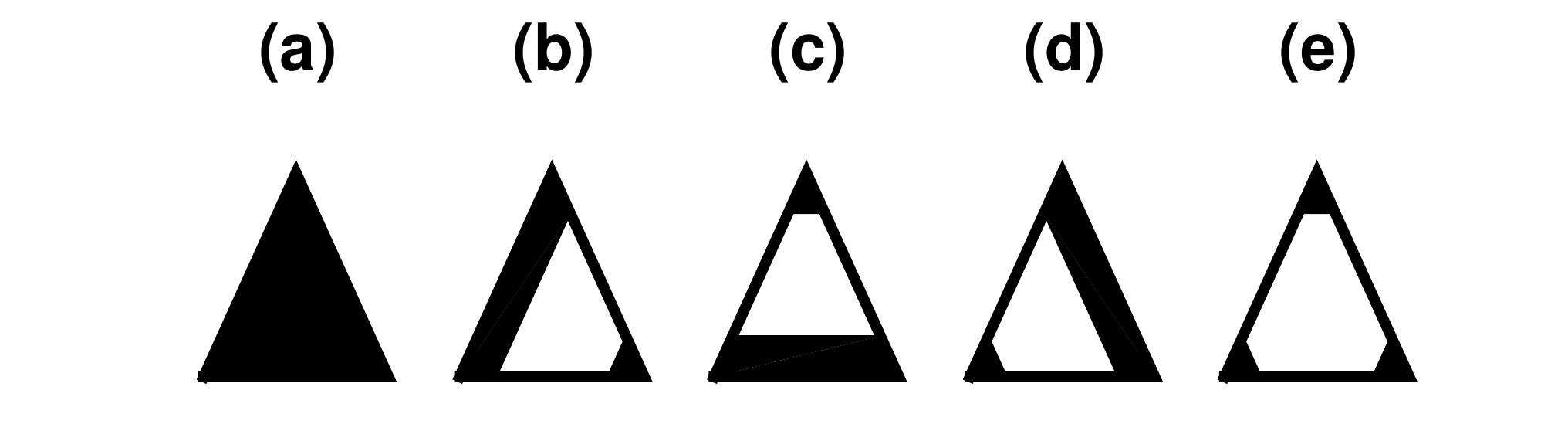}
	\caption{Schematic diagrams representing $T_n(x)$ (a), $S_n(x,y)$ (b), $S_n(x,z)$ (c), $S_n(y,x)$ (d), and $W_n(x,y,z)$ (e). With $x,y$ and $z$ we indicate the conjugated variables of the components connected with the bottom left, bottom right and top node respectively.}
	\label{fig4}
\end{figure}
\begin{figure}
    \includegraphics[width=0.99\columnwidth]{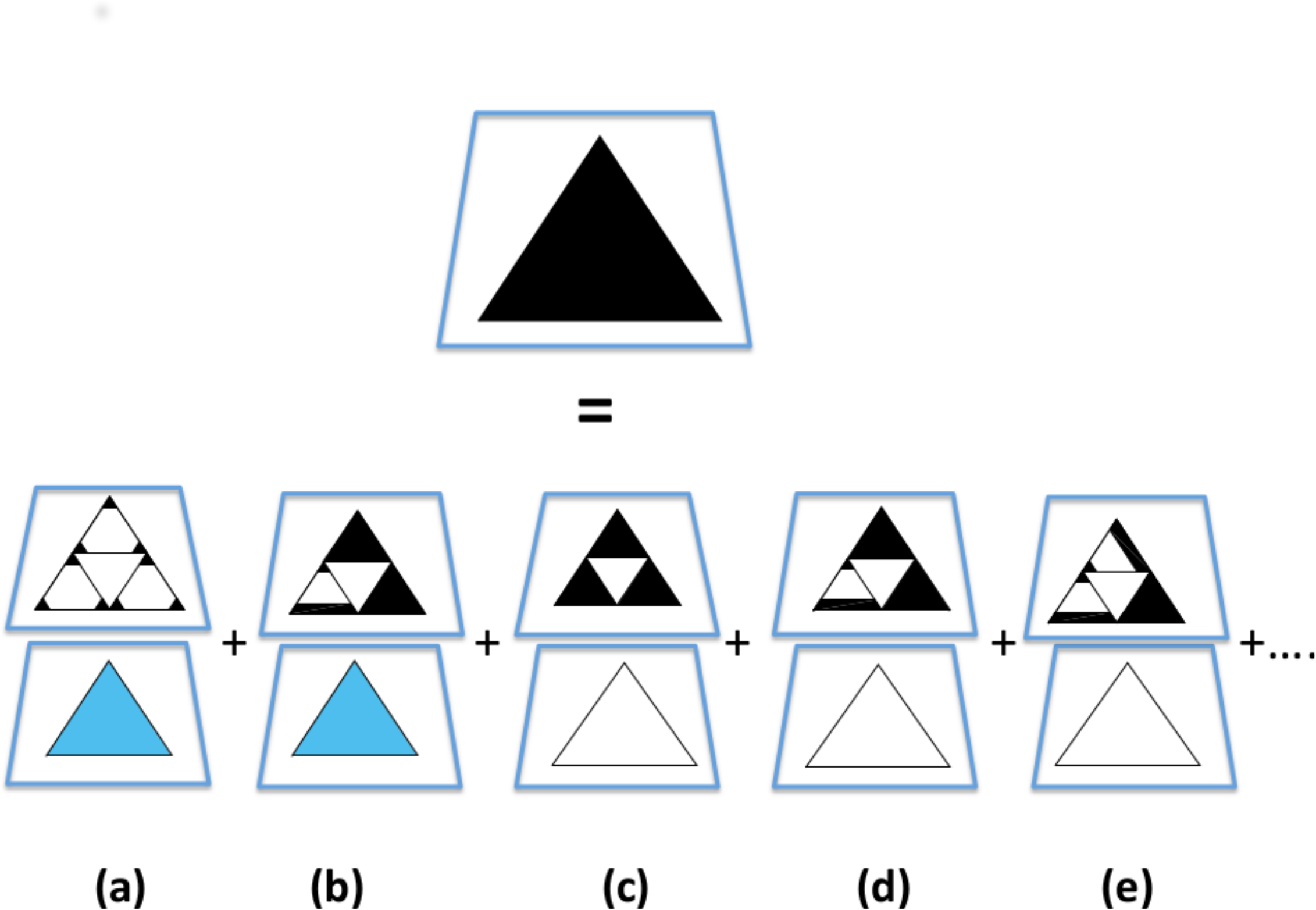}
	\caption{Diagrammatic expression for {some terms of } the recursive equation for $T_{n+1}(x)$ {given in the Appendix}. The different terms indicate $pW_n^3(x,1,1)$ (a), $pT_n^2(x)S_n(x,x)$ (b), $(1-p)T_n^3(x)$ (c), $(1-p)T_n^2(x)S_n(x,x)$ (d), and $(1-p)T_n(x)S_n(x,x)$ (e).}
	\label{fig5}
\end{figure}
\begin{figure}
    \includegraphics[width=0.99\columnwidth]{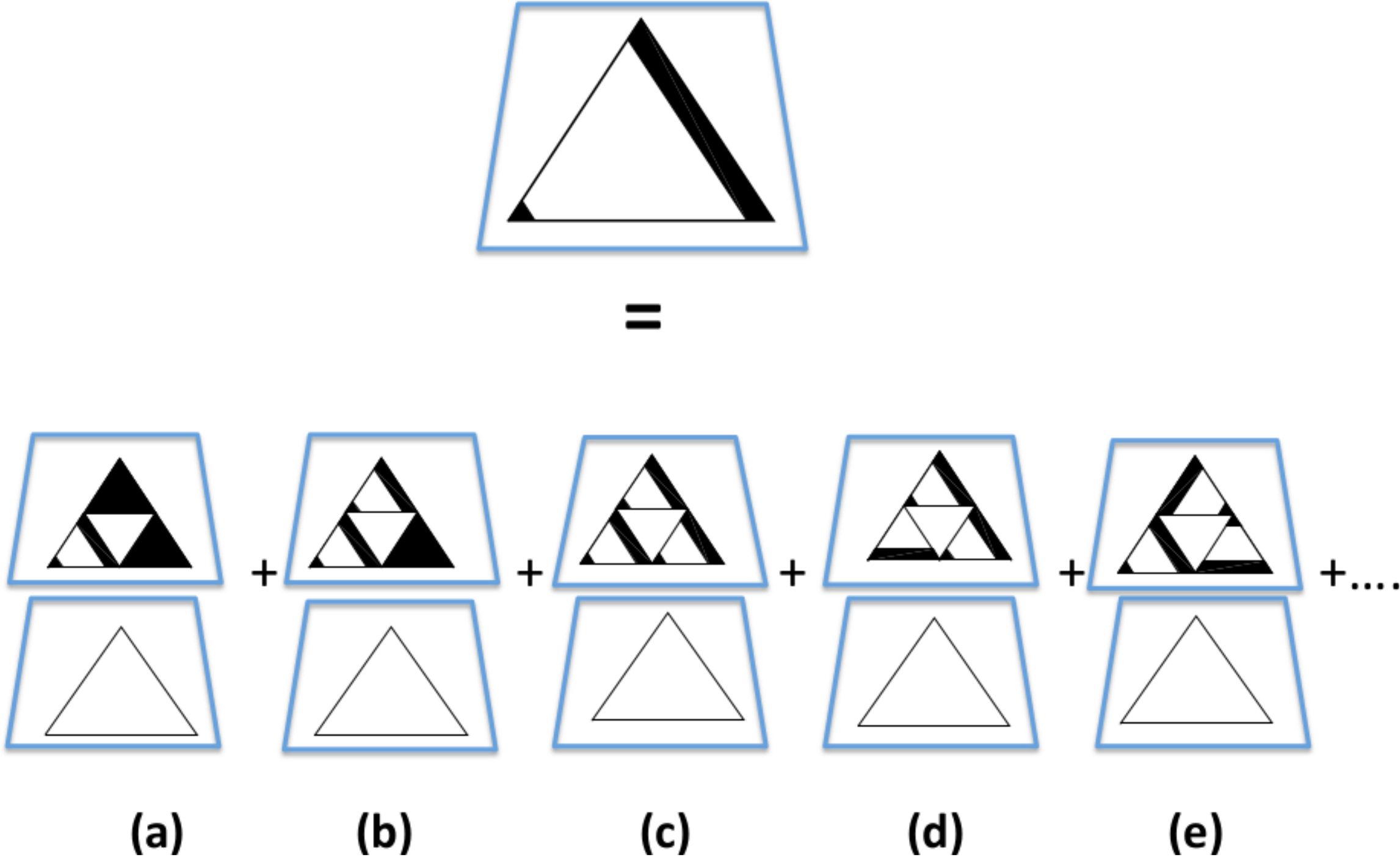}
	\caption{Diagrammatic expression  for the recursive equation for $S_{n+1}(y,x)$. The different terms indicate $(1-p)T_n^2(y)S_n(y,x)$ (a), $(1-p)T_n(y)S_n(y,y)S_n(y,x)$ (b), $(1-p)S_n(1,x)S_n^2(y,1)(x)$ (c), $(1-p)S_n(x,1)S_n(y,1)S_n(y,x)$ (d), and $(1-p)S_n(y,x)S_n^2(y,1)$ (e).}
	\label{fig6}
\end{figure}
The generating functions $T_n(x),S_n(x,y),W_n(x,y,z)$ can be expressed diagrammatically as shown in Fig.\  $\ref{fig4}$. This description is taking advantage of the relation between the multiplex Sierpinski gasket and the $d=3$ hyperbolic manifold. In particular indicating with  $x,y,z$ the conjugated  
variables to the number of links connected to the  initial three links of the $d=3$ hyperbolic manifolds Fig.\  $\ref{fig4}$(a)-(e) represent $T_n(x),S_n(x,y),S_n(x,z),S_n(y,x),W_n(x,y,z)$ respectively. In fact in this aggregated Sierpinski gasket where nodes represent the links of the original aggregated $d=3$ hyperbolic network, $T_n(x)$ is the generating function of the component connected to the three nodes of the triangle, $S_n(x,y)$ is the generating function of the two separate component connected to two out of thee nodes and to the remaining node respectively, and $W_n(x,y,z)$ is the generating function of the three separate components connected to each node of the aggregated Sierpinski gasket.

For $T_n(x),S_n(x,y)$ and $W_n(x,y,z)$ it is possible to write down recursive equations implementing the renormalization group on these structures (see Appendix).
These recursive equations can be written down by using a long but straightforward diagrammatic expansion. In Figs.\  $\ref{fig5}$ and  $\ref{fig6}$ we show {a} few terms of these diagrammatic expression for calculating $T_{n+1}(x)$ and $S_{n+1}(y,x)$. The terms that contribute to $T_{n+1}(x)$
are of  two types : 
\begin{itemize}
\item[(i)]
The terms in which the three initial nodes (of the Sierpinski gasket) are directly connected by a triangle at iteration $n=0$ and in addition they can be connected to other nodes if one takes into account  any other iteration (see for instance Fig.\  $\ref{fig5}$(a) and Fig.\  $\ref{fig5}$(b)).  
\item[(ii)]
The terms in which the three initial nodes (of the Sierpinski gasket) are not directly connected by a triangle at iteration $n=0$ but  they are connected if one takes into account  any other iteration (see for instance Fig.\  $\ref{fig5}$(c)-Fig.\  $\ref{fig5}$(e) ).
\end{itemize}
 The  terms that contribute to $S_{n+1}(y,x)$ only include diagrams in which the three initial nodes (of the Sierpinski gasket) are not directly connected (see Fig.\  $\ref{fig5}$).

To study triangle percolation our  first aim is to calculate the average size  $R_n$ of the component  connected to each one of the three initial links, i.e.,
\bea
R_n=\left.\frac{dT_n(x)}{dx}\right|_{x=1}.
\eea
and to evaluate the fractal exponent $\psi$ characterizing its asymptotic scaling for $n\gg 1$
\bea
R_n=[N_n^{(1)}]^{\psi}.
\eea
Our second aim is to study the upper percolation transition using the order parameter  $M_n$ given by 
\bea
M_n=\frac{R_n}{N_n^{(1)}}
\eea
in the limit $n\to \infty$.
 \begin{figure*}
    \includegraphics[width=1.99\columnwidth]{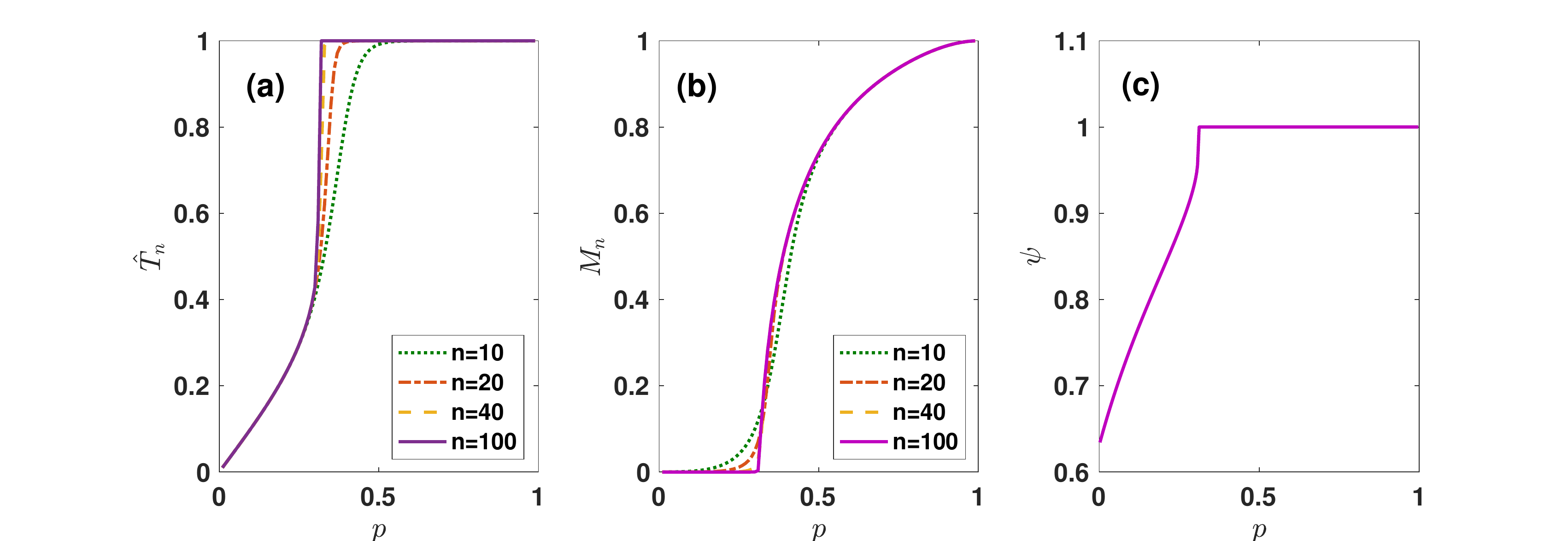}
	\caption{We plot the statistical mechanics quantities characterizing  triangle percolation in the $d=3$ hyperbolic manifold as a function of $p$: the probability $\hat{T}_{n}$ that the three initial links are  connected at iteration $n=10,20,40,100$ (panel a);  the fraction $M_n$ of links in the largest component at iterations $n=10,20,40,100$  (panel b); the fractal critical exponent $\psi$ (panel c).}
	\label{fig7}
\end{figure*}

\begin{figure}
    \includegraphics[width=0.80\columnwidth]{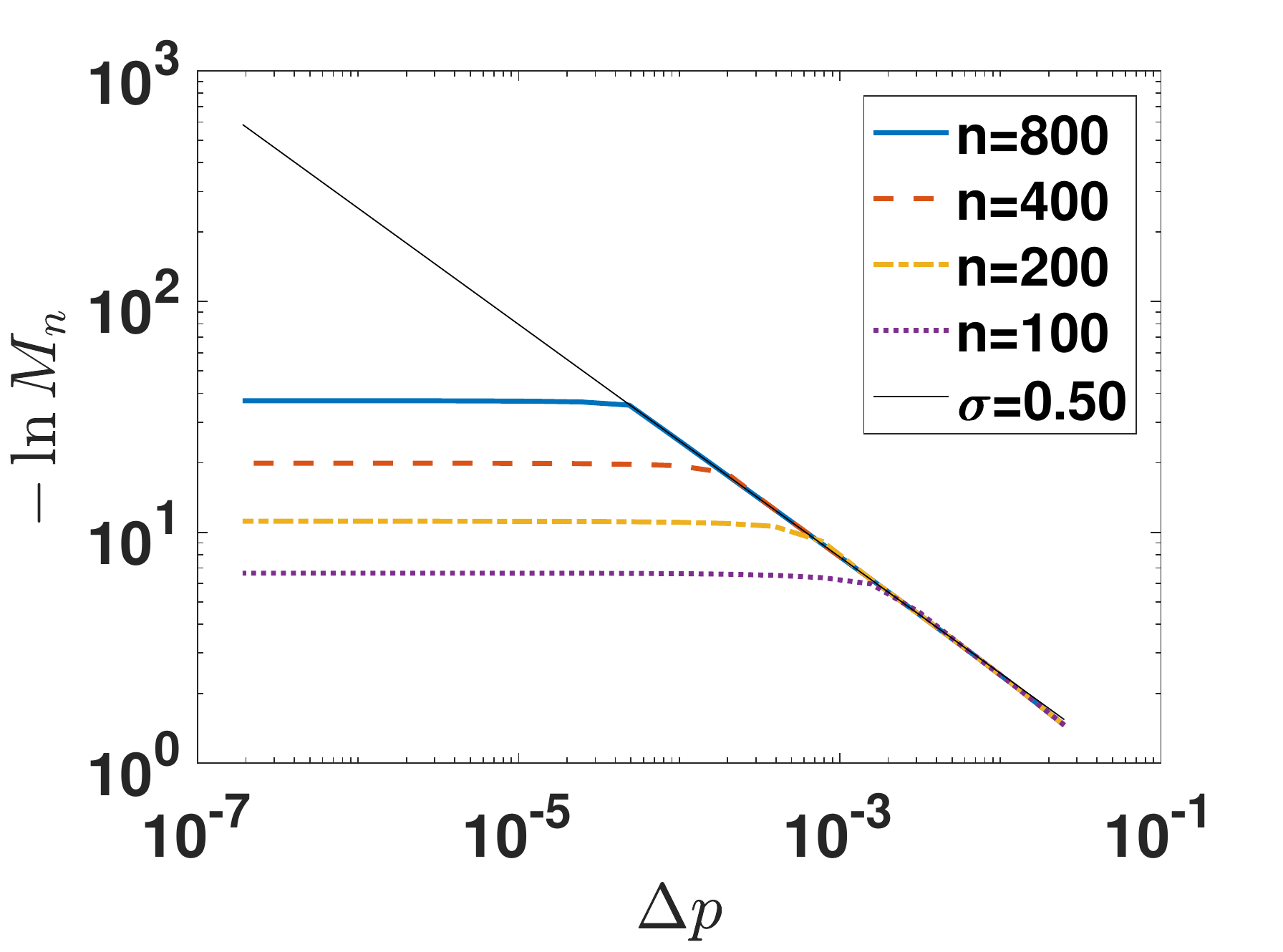}
	\caption{The BKT critical scaling of $M_n$ versus $\Delta p=p-p_c$ of triangle percolation  on the $d=3$ hyperbolic network  for $n=100$, 200, 400, 800 is plotted together with the theoretical expectation $-\ln M_{\infty}\simeq -d|\Delta p|^{-\sigma} $ with $\sigma=0.5$ for $\Delta p \ll 1$.}
	\label{fig8}
\end{figure}
To this end we note the both goals can be achieved if we characterize $T_n(x)$. From the recursive equations valid for the generating functions, we note that $T_{n+1}(x)$ depends only  on the variables $S_n(x,x),S_n(x,1),S_n(1,x),W_n(x,x,x),W_n(x,x,1)$ and $W_n(x,1,1)$. Therefore $T_n(x)$ can be found by solving a recursive nonlinear system of equations for these variables (see Appendix and Supplemental Material \cite{SI}).
If we define  vector ${\bf V}_n(x)$ whose components are given by 
\bea
&{{V}}_n^1(x)=T_n(x),&\ V_n^2(x)=S_n(x,x),\nonumber \\
&V_n^3(x)=S_n(x,1),&\ V_n^4(x)=S_n(1,x),\nonumber \\
&V_n^5(x)=W_n(x,x,x),&\ V_n^6(x)=W_n(x,x,1),\nonumber \\
&V_n^7(x)=W_n(x,1,1).&
\eea
this system of equations can be written as a recursive equation for ${\bf V}_{n+1}(x)$ given by 
\bea
{\bf{V}}_{n+1}(x)={\bf F}({\bf V}_n(x),x).
\label{Veq}
\eea 
Finally this system of equations can be differentiated obtaining
\bea
\frac{d{\bf V}_n(x)}{dx}=\sum_{i=1}^7\frac{\partial{\bf F}(x)}{\partial{V}_n^{i}(x)}\frac{d{ V}_n^{i}(x)}{dx}+\frac{\partial{\bf F}(x)}{\partial x},
\label{Vprime}
\eea
with initial condition ${\bf V}^{\prime}=(0,0,0,0,0,0,0)$ (the initial nodes are not counted).
Since the non-homogeneous term ${\partial{\bf F}(x)}/{\partial x}$  is subleading with respect to the homogeneous one,  for $n\gg 1$ 
\bea
{\bf V}_n^{\prime}(x)\propto (\lambda_J)^n
\eea
with $\lambda_J$ indicating the maximum eigenvalue of the matrix
\bea
J_{ij}=\left.\frac{\partial{F}^{i}(x)}{\partial{V}^{j}(x)}\right|_{{\bf V}(x)={\bf V}_{\infty}(1);x=1}.
\eea
This implies that $R_n$ for $n\gg1 $ scales like
\bea
R_n\sim [N_n^{(1)}]^{\psi}
\eea
with  the fractal critical exponent $\psi$  given by
\bea
\psi=\frac{\ln \lambda_J}{\ln 3}.
\eea
This fractal critical exponent $\psi$ is plotted in Fig.\  $\ref{fig7}$ where one can note its discontinuity at $p=p_2^{u}$.
Finally  by  evaluating recursively the system of Eqs.\ ($\ref{Vprime}$) we can calculate
\bea
M_n=\frac{1}{N_n^{(1)}}\left.\frac{d{V}_n^1(x)}{dx}\right|_{x=1}.
\eea
The dependence of $M_n$ on $p$  is plotted in Fig.\  $\ref{fig7}$ for increasing values of $n$ from which we can observe a very sharp but continuous transition at $p_2^{u}$.
A careful finite size analysis (see \cite{SI}) of the critical behavior of $M_n$ for $\Delta p\ll 1$ (see Fig.\  $\ref{fig8}$) reveals that the BKT nature of the  transition at $p_2^{u}$ with a critical scaling  valid for $\Delta p \ll 1$
\bea
M_{\infty}=Ae^{-d|\Delta p|^{-\sigma}}
\label{MB2}
\eea
with $A,d$ positive constants and $\sigma=0.50$.
This behavior is in agreement with renormalization-group \cite{renormalization} general results on hierarchical  networks according to which a hybrid transition in $\hat{T}_{\infty}$ should result in BKT transition for $M_{\infty}$.
Note the BKT transition has been also observed for percolation  \cite{BKT2} and for the Ising model \cite{Ising_BKT,Berker_BKT} in other hierarchical networks.

\subsection*{(B3) Tetrahedron percolation} In tetrahedron percolation, tetrahedra are removed with probability $q$ and triangles are connected to triangles through intact tetrahedra.
This is node percolation on the  Cayley tree network of degree $z=4$ where nodes are tetrahedra and links connect two adjacent tetrahedra along a triangular face. 
By indicating with $R_n$ the average number of tetrahedra connected to the initial triangle at iteration $n$ we have $R_1=p$ and 
\bea
R_n=p(1-z)R_{n-1}
\label{RnTetrahedron}
\eea
{(identical to Eqs. (\ref{RnTriangle}) and (\ref{RnUpperLink}))} with explicit solution 
\bea
R_n=[p(1-z)]^{n-1}R_1.
\eea 
Therefore, for $n\gg1$, $R_n$ scales as
\bea
R_n\simeq [N_n^{(3)}]^{\psi}
\label{scalingt3}
\eea
where the fractal critical exponent is given by 
 \bea
\psi=\frac{\ln[p(z-1)]}{\ln 3}=\frac{\ln[3p]}{\ln 3}.
\label{psit3a}
\eea
{taking $z = 4$}.  Therefore we find 
that the lower and upper percolation thresholds are given by 
\bea
p_3^{l}=\frac{1}{3}, \ p_4^{u}=1.
\eea
The order parameter $M_{\infty}$ given by 
\bea
M_{\infty}=\lim_{n\to \infty}\frac{R_n}{N^{(3)}_n}
\label{Minft3a}
\eea
and indicating the fraction  of tetrahedra in the largest component, has 
 a discontinuous critical behavior,  i.e.,
 \bea
M_{\infty}&=&\left\{\begin{array}{ccc}0 &\mbox{if} &p<p_3^{u}\\
1 &\mbox{if}&p=p_3^{u}\end{array}\right..
\label{MB3}
\eea

\begin{figure}
    \includegraphics[width=0.99\columnwidth]{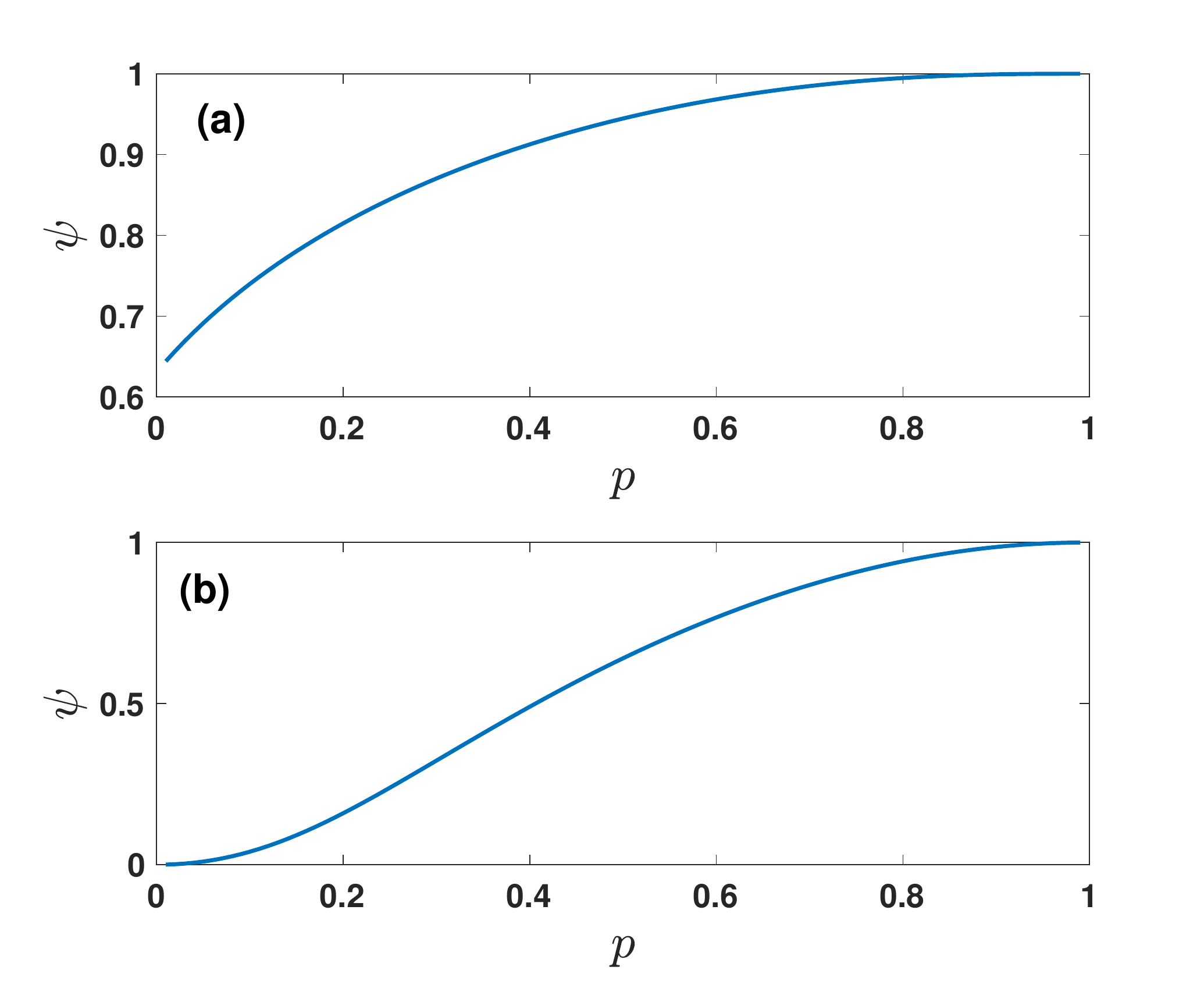}
	\caption{The fractal critical  exponent ${\psi}$ plotted versus $p$ for node percolation (panel a) and upper-link percolation (panel b) on the $d=3$ hyperbolic manifold. }
	\label{Psi_nodes_links}
\end{figure}

\subsection*{(B4)  Node percolation} In node percolation nodes are removed with probability $q$ and links are connected to links through intact nodes.\\

To study this percolation problem  we consider the variables ${R}_n^{[+++]}$, ${R}^{[++-]}_n$ and ${R}_n^{[+--]}$ indicating (at iteration $n$) the average number of nodes of the component connected to the initial triangle that has  three intact nodes, only two intact nodes, or only {one}   intact node, respectively  (cases $[+++],[++-],[+--]$).
Starting from the initial condition ${R}^{[+++]}_0={R}^{[++-]}_0=R^{[+--]}_0=0$ the values of ${R}^{[+++]}_n,{R}^{[++-]}_n $ and ${R}^{[+--]}_n$ are found by iteration of the equations, 
\bea
{\bf R}_{n+1}={\bf B}{\bf R}_n+p{\bf 1}
\label{B}
\eea
where 
\bea
{\bf R}_n=\left(\begin{array}{c}{R}_{n}^{[+++]}\\{R}_{n}^{[++-]}\\{R}_{n}^{[+--]}\end{array}\right), \  \ {\bf 1}=\left(\begin{array}{c}1\\1\\1\end{array}\right)
\eea
and 
\bea
{\bf B}=\left(\begin{array}{ccc} 3p& 3(1-p)&0\\p&2p+(1-p)& 2(1-p)\\0& 2p&p+2(1-p)\end{array}\right).
\eea
The solution of Eq.\  (\ref{B}) is given by 
\bea
{\bf R}_n=p\sum_{r=0}^{n-1}{\bf B}^r{\bf 1}.
\eea
Therefore by indicating with  ${\lambda}$  the maximum eigenvalue of ${\bf B}$, the leading term for $n\gg 1$ goes like
\bea
{\bf R}_n\sim p{\lambda}^n \sim  [N_n^{(0)}]^{{\psi}}
\eea
with 
\bea
{\psi}=\frac{\ln {\lambda}}{\ln 3}.
\eea
In Fig. $\ref{Psi_nodes_links}$(a) we plot ${\psi}$ versus $p$.

By imposing that the maximum eigenvalue $\psi=1$ we get $p_4^{u}$ (extensive cluster) {and by} imposing $\psi=0$ we get $p_4^{l}$.
The percolation thresholds are therefore found at 
\bea
p_4^{l}=0,\ p_4^{u}=1.
\eea
The order parameter $M_{\infty}$ given by 
\bea
M_{\infty}=\lim_{n\to \infty}\frac{R_n}{N^{(0)}_n}
\label{Minft3}
\eea
and indicating the fraction of nodes in the largest component, has 
 a discontinuous critical behavior  i.e.,
 \bea
M_{\infty}&=&\left\{\begin{array}{ccc}0 &\mbox{if} &p<p_4^{u}\\
1 &\mbox{if}&p=p_4^{u}\end{array}\right..
\label{MB4}
\eea

\subsection*{(B5) Upper-link percolation} 
In upper-link percolation links are removed with probability $q$ and triangles are connected to triangles through intact links.

 In order to characterize this topological percolation problem we consider the variables ${R}_n^{[+++]}$, ${R}^{[++-]}_n$ and ${R}_n^{[+--]}$ indicating  (at iteration $n$)  the average number of links of the component connected to an intact link of an initial triangle  given that the initial triangle has three intact links, only two intact links, or only one intact link respectively  (cases $[+++],[++-],[+--]$).
Starting from the initial condition ${R}^{[+++]}_0={R}^{[++-]}_0={R}^{[+--]}_0=0$ the values of ${R}^{[+++]}_n,{R}^{[++-]}_n $ and ${R}^{[+--]}_n$ are found by iteration of the equations,
\bea
{\bf R}_{n+1}={\bf B}{\bf R}_n+{\bf R}_1
\label{B}
\eea
where 
\bea
{\bf R}_n=\left(\begin{array}{c}{R}_{n}^{[+++]}\\{R}_{n}^{[++-]}\\{R}_{n}^{[+--]}\end{array}\right), \  \ {\bf R}_1=(p^3+6p^2q+2pq^2)\left(\begin{array}{c}1\\ 1 \\1\end{array}\right)\nonumber 
\eea
and 
%\begin{widetext}
\bea
\hspace*{-6mm}{\bf B}=\left(\begin{array}{ccc} 3p^3+3p^2q& 4p^2q+6pq^2&pq^2+q^3\\2p^3+2p^2q&p^3+5p^2q+3pq^2& 2p^2q+pq^2+q^3\\p^3+p^2q&2p^3+2p^2q+2pq^2&2p^2q+pq^2+q^3\end{array}\right)\   \cr
\label{Bul}
\eea
where $q=1-p$.
%\end{widetext}
The solution of Eq.\  (\ref{B}) is given by 
\bea
{\bf R}_n=\sum_{r=0}^{n-1}{\bf B}^r{\bf R}_1.
\eea
Therefore by indicating with  ${\lambda}$  the maximum eigenvalue of ${\bf B}$ given by Eq.\ ($\ref{Bul}$), the leading term  for $n\gg 1$ goes like
\bea
{\bf R}_n\sim p{\lambda}^n\sim  [N_n^{(1)}]^{{\psi}}
\eea
with 
\bea
{\psi}=\frac{\ln {\lambda}}{\ln 3}. 
\eea
 The  numerically evaluated the fractal exponent $\psi$ is plotted versus $p$ in
 Fig.\  $\ref{Psi_nodes_links}$b.
By imposing $\psi=0$ we get $p_5^{l}$, by imposing $\psi=1$ we get $p_5^{u}$.
The percolation threshold are therefore found at 
\bea
p_5^{l}=0,\ p_5^{u}=1.
\eea
\\
The order parameter $M_{\infty}$ is  given by 
\bea
M_{\infty}=\lim_{n\to \infty}\frac{R_n}{N^{(1)}_n}
\label{Minft3}
\eea
and indicating the fraction of links in the largest component has 
 a discontinuous critical behavior, i.e.,
 \bea
M_{\infty}&=&\left\{\begin{array}{ccc}0 &\mbox{if} &p<p_5^{u}\\
1 &\mbox{if}&p=p_5^{u}\end{array}\right.. \\
\label{MB5}
\eea
\subsection*{(B6) Upper-triangle percolation}

In upper-triangle percolation triangles are removed with probability $q$ and  tetrahedra are connected to tetrahedra through intact triangles.
This is link percolation on the  Cayley tree network of degree $z=4$ where nodes are triangles and links connect two adjacent triangles. \\
By indicating with $R_n$ the average number of tetrahedra connected to the initial triangle at iteration $n$ we have $R_1=p$ and for $n>1$ we have 
\bea
R_n=p(1-z)R_{n-1}.
\eea
These are the same equations (Eqs. (\ref{RnTetrahedron})) found in  tetrahedron  percolation. 
Therefore by  taking $z=4$, we find 
that the lower and upper percolation thresholds are given 
\bea
p_6^{l}=\frac{1}{3}, \ p_6^{u}=1
\eea
 with the same critical behavior found for tetrahedron percolation. 
Therefore $R_n$ for $n\gg1 $ obeys the scaling in Eq.\  $(\ref{scalingt3})$ with  the fractal critical exponent $\psi$ given by Eq.\  (\ref{psit3a}).
 Moreover at $p_6^{u}$ the order parameter $M_{\infty}$ defined as in Eq.\  (\ref{Minft3a}) has a discontinuous jump described by 
\bea
M_{\infty}&=&\left\{\begin{array}{ccc}0 &\mbox{if} &p<p_6^{u} \\
1 &\mbox{if}&p=p_6^{u}\end{array}\right..\\
\label{MB6}
\eea  

\section{Comparison between topological percolation in $d=2$ and $d=3$ hyperbolic simplicial complexes}
Our detailed study of  topological percolation in the $d=2$ and $d=3$ hyperbolic simplicial complexes (performed in Secs.\ V and VI) can be summarized by considering the following major points.

Topological percolation of the $d=2$ hyperbolic manifold displays the following major properties:
\begin{itemize}
\item[-] All topological percolation problems have two percolation thresholds $p^l$ and $p^u$ (given in Table $\ref{table1}$).
\item[-]All topological percolation problems are discontinuous at the upper critical percolation threshold $p^u$. However with the exception of the nontrivial link percolation problem (A1) studied in Ref.\ \cite{hyperbolic_Ziff} the upper critical percolation threshold is  $p^u=1$ and the discontinuity takes the form of a trivial 0-1 law for the order parameter $M_{\infty}$ (see Eqs. $(\ref{MA2}),(\ref{MA3})$ and $(\ref{MA4})$).
\end{itemize}

Topological percolation  of the $d=3$ hyperbolic manifold displays the following major properties:
\begin{itemize}
\item[-] All topological percolation problems have two percolation thresholds $p^l$ and $p^u$ (given in Table $\ref{table1}$) except link percolation (B1) studied in Ref.\ \cite{percolation_Apollonian} for which $p^u=0$ and $p^l$ cannot be defined. 
\item[-] All topological percolation problems with the exception of link percolation (B1) and triangle percolation (B2) are  discontinuous at the upper critical percolation threshold $p^u=1$ obeying the trivial $0-1$ law (see Eqs. $(\ref{MB3}),(\ref{MB4}),(\ref{MB5})$ and $\ref{MB6}$).
\item[-] Link percolation (B1) studied in Ref.\ \cite{percolation_Apollonian} is continuous at the upper percolation threshold $p^u=0$ with  an exponential critical behavior (see Eqs. $\ref{MB1}$) typical of scale-free networks.
\item[-] 
Triangle percolation (B2) is continuous at the upper percolation threshold $p^u$. Moreover it  follows an anomalous critical behavior displaying a  BKT transition (see  Eqs. $(\ref{MB2})$). We observe that the BKT transition is not seen for any other topological percolation problem in the same manifold or in the $d=2$ hyperbolic manifold.
\item[-] 
In topological percolation on the   $d=3$ hyperbolic manifold not even one topological percolation problem displays a nontrivial discontinuity, i.e., there is no percolation problem with a critical behavior similar to link percolation (A1) in the $d=2$ hyperbolic manifold.
\end{itemize}

\section{Conclusions}

In this paper we have proposed topological percolation which extends the study of percolation beyond node and link {(site and bond)} percolation for simplicial complexes.
Topological percolation on a $d$-dimensional simplex refers to a set of $2d$ percolation processes that reveal different topological properties of the simplicial complex. As such this approach can be used on any arbitrary simplicial complex dataset and has relevance for the wide variety of applications where simplicial complexes are used from brain research to social networks.
Here we have treated topological percolation on two major examples of $d=2$  and $d=3$ hyperbolic simplicial complexes  describing discrete manifolds. The $d=2$ hyperbolic manifold is the Farey simplicial complex, the $d=3$ hyperbolic simplicial complex has {a} skeleton given by the Apollonian graph.
We have emphasized  the ubiquitous presence of two percolation thresholds characteristic of hyperbolic discrete structures for each topological percolation problem {with} the exception of the trivial link percolation for the $d=3$ manifold (see Table $\ref{table1}$). However we have observed that the nature of the phase transitions can vary and that topological percolation  reveals properties  that are unexpected if one just focuses on the robustness of the network  skeleton. In particular  here we show that  triangle percolation in the $d=3$  hyperbolic manifold  displays a  BKT transition while no similar critical behavior is observed at the level of link percolation on the same simplicial complex.
Moreover we observe an important dependence of topological percolation with the dimension of the simplicial complex. In fact while the $d=2$ Farey simplicial complex and the $d=3$ hyperbolic manifold considered here obey similar generation rules, the BKT transition is only observed in topological percolation of the $d=3$ manifold. Additionally the peculiar properties of the nontrivial discontinuous link percolation on $d=2$ hyperbolic manifold are not observed for any of the six topological percolation problems in the $d=3$ manifold.

In conclusion our work constitutes one of the few studies of percolation on hyperbolic manifolds of dimension $d=3$ and shows  the rich unexpected behavior of topological percolation in higher-dimensional simplicial complexes.

{This work can be extended in different directions. The analytical investigation of topological percolation can be extended to hyperbolic cell complexes and higher-dimensional hyperbolic simplicial complexes.
Moreover topological percolation can be explored on the increasing number of available simplicial complex datasets such a collaboration networks, social networks or protein interaction networks which include an important interplay between randomness and order.}

\section*{Acknowledgements}
G.B. was partially supported by the  Perimeter Institute for Theoretical Physics (PI). The PI is supported by the Government of Canada through Industry Canada and by the Province of Ontario through the Ministry of Research and Innovation.

\newpage

\section*{Appendix: (B2) $d=3$ generating functions}
In this appendix we write the recursive equations for the generating functions solving  triangle percolation in the $d=3$ hyperbolic manifold and we write explicitly the Eqs.\ (\ref{Veq}).
Using the diagrammatic expansion discussed in the main text it is possible to derive the following recursive equations for the generating functions.  
Starting  from the initial condition $T_0(x)=p, S_0(x,y)=0$ and $W_0(x,y,z)=1-p$ (the initial nodes are not counted) the recursive equations for $T_n(x), S_n(x,y) $ and $W_n(x,y,z)$ read
\begin{widetext}
\bea
T_{n+1}(x)&=&p\left\{x^3T_n^3(x)+9x^2T^2_{n}(x)S_n(x,x)+3x^3T^2_n(x)W(x,x,x)
+24x^3T_n(x)S_n^2(x,x)+3x^2T_n(x)S^2_n(x,1)\right.\nonumber \\
&& \left. +12x^3T_n(x)S_n(x,x)W_n(x,x,x)+6x^2T_n(x)S_n(x,1)W_n(x,x,1)+3x^2T_n(x)W_n^2(x,x,1)\right.\nonumber \\
&&\left.+14x^3 S_n^3(x,x)+9x^2S_n(x,x)S_n^2(x,1)+S_n^3(1,x)\right.\nonumber \\
&&\left.+3xS^2_n(x,1)S_n(1,x)+3S^2_n(1,x)W_n(x,1,1)+6xS_n(x,1)S_n(1,x)W_n(x,x,1)\right.\nonumber \\
&&\left. +12x^2S_n(x,1)S_n(x,x)W_n(x,x,1)+3xS_n^2(x,1)W(x,1,1)\right.\nonumber \\
&&\left.+3x^2S^2_n(x,1)W(x,x,x)+3S_n(1,x)W^2_n(x,1,1)+6S_n(x,1)W_n(x,1,1)W_n(x,x,1)+W^3_n(x,1,1)\right\}\nonumber \\
&&+(1-p)\left\{x^3T_n^3(x)+6x^3T^2_n(x)S(x,x)+3x^2T_n(x)S_n^2(x,1)\right\}\nonumber \\
S_{n+1}(x,y)&=&(1-p)\left\{x^3T_n^2(x)S_n(x,y)+x^3T^2_n(x)W_n(x,x,y)+4x^3T_n(x)S_n(x,x)S_n(x,y)\right.\nonumber \\
&&\left.+2x^2yT_n(x)S_n(y,x)S_n(x,y)+2x^2T_n(x)S_n(x,1)W_n(x,y,1)+xy^2T_n(y)S_n^2(x,y)\right.\nonumber \\
&&\left.+xS^2_n(x,1)S_n(1,y)+2xy S_n(x,1)S_n(x,y)S_n(y,1)+xS_n^2(x,1)W_n(y,1,1)+x^2S^2_n(x,1)S_n(x,y)\right\}\nonumber \\
W_{n+1}(x,y,z)&=&(1-p)\left\{x^3T_n(x)S_n(x,y)S_n(x,z)+x^3T_n(x)S_n(x,z)W_n(x,x,y)+x^2zT_n(x)S_n(z,x)W_n(x,y,z)\right.\nonumber \\&&+x^3T_n(x)S_n(x,y)W_n(x,x,z)+x^2yT_n(x)S_n(y,x)W_n(x,y,z)+x^2T_n(x)W_n(x,y,1)W_n(x,z,1)\nonumber \\
&&+z^3T_n(z)S_n(z,x)S_n(z,y)+z^3T_n(z)S_n(z,x)W_n(y,z,z)+xz^2T_n(z)S_n(x,z)W_n(x,y,z)\nonumber \\&&+z^3T_n(z)S_n(z,y)W_n(x,z,z)+yz^2T_n(z)S_n(y,z)W_n(x,y,z)+z^2T_n(z)W_n(x,z,1)W_n(y,z,1)\nonumber \\
&&+y^3T_n(y)S_n(y,x)S_n(y,z)+y^3T_n(y)S_n(y,x)W_n(y,y,z)+xy^2T_n(y)S_n(x,y)W_n(x,y,z)\nonumber \\
&&+y^3T_n(y)S_n(y,z)W_n(x,y,y)+y^2zT_n(y)S_n(z,y)W_n(x,y,z)+y^2T_n(y)W_n(x,y,1)W_n(z,y,1)\nonumber \\
&&+S_n(1,x)S_n(1,y)S_n(1,z)+2y^3S_n(y,x)S_n(y,y)S_n(y,z)+S_n(1,x)S_n(1,z)W_n(y,1,1)\nonumber\\&&+z^3S_n(z,x)S_n(z,z)S_n(z,y)+yz^2S_n(z,x)S_n(z,y)S_n(y,z)+z^2S_n(z,x)S_n(z,1)W_n(y,z,1)\nonumber\\&&+z^3S_n(z,x)S_n(z,z)S_n(z,y)+y^2zS_n(y,x)S_n(z,y)S_n(y,z)+zS_n(1,x)S_n(z,1)W_n(y,z,1)\nonumber\\
&&+S_n(1,x)S_n(1,y)W_n(z,1,1)+yS_n(1,x)S_n(y,1)W_n(y,z,1)+y^2S_n(y,x)S_n(y,1)W_n(y,z,1)\nonumber \\
&&+S_n(1,x)W_n(y,1,1)W_n(z,1,1)+x^3S_n(x,x)S_n(x,z)S_n(x,y)+x^2yS_n(x,y)S_n(y,x)S_n(x,z)\nonumber \\
&&+x^2S_n(x,1)S_n(x,z)W_n(x,y,1)+xz^2S_n(x,z)S_n(z,x)S_n(z,y)+xyzS_n(x,y)S_n(z,x)S_n(y,z)\nonumber \\&&+xzS_n(x,1)S_n(z,x)W_n(y,z,1)+xS_n(x,1)S_n(1,y)W_n(x,z,1)+xyS_n(x,1)S_n(y,1)W_n(x,y,z)\nonumber\\&&+xyS_n(x,y)S_n(y,1)W_n(x,z,1)+xS_n(x,1)W_n(x,z,1)W_n(y,1,1)\nonumber\\
&&+x^3S_n(x,x)S_n(x,y)S_n(x,z)+xy^2S_n(x,y)S_n(y,x)S_n(y,z)+xS_n(x,1)S_n(1,z)W_n(x,y,1)\nonumber\\
&&+x^2zS_n(x,z)S_n(x,y)S_n(z,x)+xyzS_n(x,z)S_n(y,x)S_n(z,y)+xzS_n(x,z)S_n(z,1)W_n(x,y,1)\nonumber\\
&&+xzS_n(x,1)S_n(z,1)W_n(x,y,z)+x^2S_n(x,1)S_n(x,y)W_n(x,z,1)+xyS_n(x,1)S_n(y,x)W_n(y,z,1)\nonumber\\
&&+xS_n(x,1)W_n(x,y,1)W_n(z,1,1)\nonumber \\
&&+S_n(1,y)S_n(1,z)W_n(x,1,1)+y^2S_n(y,z)S_n(y,1)W_n(x,y,1)+yS_n(y,1)S_n(1,z)W_n(x,y,1)\nonumber \\
&&+S_n(1,z)W_n(x,1,1)W_n(y,1,1)+zS_n(1,y)S_n(z,1)W_n(x,z,1)+yzS_n(z,y)S_n(y,1)W_n(x,z,1)\nonumber \\
&&+yzS_n(z,1)S_n(y,1)W_n(x,y,z)+zS_n(z,1)W_n(x,z,1)W_n(y,1,1)+z^2S_n(z,1)S_n(z,y)W_n(x,z,1)\nonumber \\
&&+yzS_n(z,1)S_n(y,z)W_n(x,y,1)+zS_n(z,1)W_n(x,1,1)W_n(y,z,1)+S_n(1,y)W_n(x,1,1)W_n(z,1,1)\nonumber \\
&&+yS_n(y,1)W_n(x,1,1)W_n(y,z,1)+yS_n(y,1)W_n(x,y,1)W_n(z,1,1)+W_n(x,1,1)W_n(y,1,1)W_n(z,1,1)\nonumber
\eea
\end{widetext}
We note that $T_{n+1}(x)$ depends only  on the variables $S_n(x,x),S_n(x,1),S_n(1,x),W_n(x,x,x),W_n(x,x,1)$ and $W_n(x,1,1)$. Therefore $T_n(x)$ can be found \cite{SI} by solving the following  recursive non-linear system of equations for these variables refered in the text as Eq.\ (\ref{Veq}),
\begin{widetext}
\bea
T_{n+1}(x)&=&p\left\{x^3T_n^3(x)+9x^2T^2_{n}(x)S_n(x,x)+3x^3T^2_n(x)W(x,x,x)
+24x^3T_n(x)S_n^2(x,x)+3x^2T_n(x)S^2_n(x,1)\right.\nonumber \\
&& \left. +12x^3T_n(x)S_n(x,x)W_n(x,x,x)+6x^2T_n(x)S_n(x,1)W_n(x,x,1)+3x^2T_n(x)W_n^2(x,x,1)\right.\nonumber \\
&&\left.+14x^3 S_n^3(x,x)+9x^2S_n(x,x)S_n^2(x,1)+S_n^3(1,x)\right.\nonumber \\
&&\left.+3xS^2_n(x,1)S_n(1,x)+3S^2_n(1,x)W_n(x,1,1)+6xS_n(x,1)S_n(1,x)W_n(x,x,1)\right.\nonumber \\
&&\left. +12x^2S_n(x,1)S_n(x,x)W_n(x,x,1)+3xS_n^2(x,1)W(x,1,1)\right.\nonumber \\
&&\left.+3x^2S^2_n(x,1)W(x,x,x)+3S_n(1,x)W^2_n(x,1,1)+6S_n(x,1)W_n(x,1,1)W_n(x,x,1)+W^3_n(x,1,1)\right\}\nonumber \\
S_n(x,x)&=&(1-p) \left\{2 x^2 T_n(x) S_n(x,1) W_n(x,x,1)+x^3 T_n^2(x) S_n(x,x)+7 x^3 T_n(x)
   S_n^2(x,x)\right.\nonumber \\
   &&\left.+x S_n^2(x,1) W_n(x,1,1)+3 x^2 S_n^2(x,1) S_n(x,x)+x S_n(1,x)
   S_n^2(x,1)+x^3 T_n^2(x) W_n(x,x,x)\right\}\nonumber \\
   S_n(x,1)&=&(1-p) \left\{2 x^2 T_n(x) S_n(x,1) W_n(x,1,1)+x^3 T_n^2(x) S_n(x,1)+4 x^3 T_n(x)
   S_n(x,1) S_n(x,x)\right. \nonumber \\
   &&\left.+2 x^2 T_n(x) S_n(1,x) S_n(x,1)+x T_n(1) S_n^2(x,1)+x
   W_n(1,1,1) S_n^2(x,1)+x^2 S_n^3(x,1)\right.\nonumber \\
   &&\left.+3 x S_n(1,1) S_n^2(x,1)+x^3T_n^2(x) W_n(x,x,1)\right\}\nonumber \\
  S_n(1,x)&=& (1-p) \left\{2 T_n(1) S_n(1,1) W_n(1,x,1)+x^2 T_n(x) S_n^2(1,x)+4 T_n(1) S_n(1,1)
   S_n(1,x)+T_n^2(1) S_n(1,x)+\right.\nonumber \\
   &&\left. 2 x T_n(1) S_n(1,x) S_n(x,1)+S_n^2(1,1)
   W_n(x,1,1)+2 S_n^2(1,1) S_n(1,x)+2 x S_n(1,1) S_n(1,x) S_n(x,1)\right.\nonumber \\
   &&\left.+T_n^2(1) W_n(1,1,x)\right\}\nonumber \\
   W_n(x,x,x)&=&(1-p) \left\{12 x^3 T_n(x) S_n(x,x) W_n(x,x,x)+3 x^3 T_n(x) S_n^2(x,x)+11 x^2
   S_n(x,1) S_n(x,x) W_n(x,x,1)\right.\nonumber \\&&\left.+4 x^2 S_n^2(x,1) W_n(x,x,x)+6 x S_n(1,x) S_n(x,1)  W_n(x,x,1)+6 x S_n(x,1) W_n(x,1,1) W_n(x,x,1)\right.\nonumber \\
   &&\left.+3 S_n(1,x) W_n^2(x,1,1)+3  S_n^2(1,x) W_n(x,1,1)+14 x^3 S_n^3(x,x)+S_n^3(1,x)+3 x^2 T_n(x)
   W_n^2(x,x,1)\right.\nonumber \\
   &&\left.+W_n^3(x,1,1)\right\}\nonumber \\
   W_n(x,x,1)&=&(1-p) \left\{4 x^3 T_n(x) S_n(x,x) W_n(x,x,1)+2 x^3 T_n(x) S_n(x,1) W_n(x,x,x)+2 x^2
   T_n(x) S_n(1,x) W_n(x,x,1)\right.\nonumber \\&&\left.+2 x T_n(1) S_n(x,1) W_n(x,x,1)+2 T_n(1) S_n(1,x)   W_n(x,1,1)+2 x^3 T_n(x) S_n(x,1) S_n(x,x)\right.\nonumber \\&&\left.+T_n(1) S_n^2(1,x)+3 x^2 S_n(x,1)S_n(x,x) W_n(x,1,1)+3 x^2 S_n^2(x,1) W_n(x,x,1)\right.\nonumber \\&&\left.+x^2 S_n(x,1) S_n(x,x) W_n(x,x,1)+2 x S_n(x,1) W_n^2(x,1,1)+4 x S_n(1,x) S_n(x,1) W_n(x,1,1)\right.\nonumber \\&&\left.+6 x S_n(1,1) S_n(x,1) W_n(x,x,1)+2 x W_n(1,1,1) S_n(x,1) W_n(x,x,1)+3 S_n(1,1)
   W_n^2(x,1,1)\right.\nonumber \\&&\left.+W_n(1,1,1) S_n^2(1,x)+6 S_n(1,1) S_n(1,x) W_n(x,1,1)+2
   W_n(1,1,1) S_n(1,x) W_n(x,1,1)\right.\nonumber \\&&\left.+6 x^3 S_n(x,1) S_n^2(x,x)+4 x^2 S_n(1,x)S_n(x,1) S_n(x,x)+2 x S_n^2(1,x) S_n(x,1)\right.\nonumber \\&&\left.+3 S_n(1,1) S_n^2(1,x)+2 x^2 T_n(x) W_n(x,1,1) W_n(x,x,1)+T_n(1) W_n^2(x,1,1)+\right.\nonumber \\&&\left. W_n(1,1,1) W_n^2(x,1,1)\right\}\nonumber \\   
   W_n(x,1,1)&=&(1-p) \left\{2 x^3 T_n(x) S_n(x,1) W_n(x,x,1)+2 x^2 T_n(x) S_n(1,x) W_n(x,1,1)+2 x
   T_n(1) S_n(x,1) W_n(x,1,1)\right.\nonumber \\&&\left.+2 T_n(1) W_n(1,1,1) S_n(1,x)+4 T_n(1) S_n(1,1)
   W_n(x,1,1)+x^3 T_n(x) S_n^2(x,1)\right.\nonumber \\&&\left.+2 T_n(1) S_n(1,1) S_n(1,x)+x^2 S_n^2(x,1) W_n(x,1,1)+x^2 S_n(x,1) S_n(x,x) W_n(x,1,1)\right.\nonumber \\&&\left.+x W_n(1,1,1) S_n(1,x) S_n(x,1)+7 x S_n(1,1) S_n(x,1) W_n(x,1,1)+2 x W_n(1,1,1) S_n(x,1) W_n(x,1,1)\right.\nonumber \\&&\left.+W_n^2(1,1,1) S_n(1,x)+6 S_n(1,1) W_n(1,1,1) S_n(1,x)+8 S_n^2(1,1) W_n(x,1,1)\right.\nonumber \\&&\left.+6 S_n(1,1)  W_n(1,1,1) W_n(x,1,1)+2 x^3 S_n^2(x,1) S_n(x,x)+2 x^2 S_n(1,x) S_n^2(x,1)\right.\nonumber \\&&\left.+4x S_n(1,1) S_n(1,x) S_n(x,1)+7 S_n^2(1,1) S_n(1,x)+x^2 T_n(x) W_n^2(x,1,1)\right.\nonumber \\&&\left.+2 T_n(1) W_n(1,1,1) W_n(x,1,1)+W_n^2(1,1,1) W_n(x,1,1)\right\}
\eea
\end{widetext}
\end{document}